\pdfoutput=1

\documentclass[cernpreprint,texlive=2016,txfonts,UKenglish]{latex/atlasdoc}

\usepackage[biblatex=true]{latex/atlaspackage}
\usepackage{latex/atlasphysics}

\newcommand{\AtlasCoordFootnote}{
ATLAS uses a right-handed coordinate system with its origin at the
nominal interaction point (IP) in the centre of the detector and the
$z$-axis along the beam pipe. The $x$-axis points from the IP to the
centre of the LHC ring, and the $y$-axis points upwards. Cylindrical
coordinates $(r,\phi)$ are used in the transverse plane, $\phi$ being
the azimuthal angle around the $z$-axis. The pseudorapidity is defined
in terms of the polar angle $\theta$ as $\eta=-\ln\tan(\theta/2)$. The
angular distance is measured in units of 
$\Delta R\equiv\sqrt{(\Delta\eta)^{2}+(\Delta\phi)^{2}}$. The rapidity
is  defined as $y=0.5\ln[(E+p_z)/(E-p_z)]$, where $E$ is the energy
and $p_z$ is the $z$-component of the momentum, and transverse energy
is defined as $E_{\mathrm T}=E\sin\theta$.}

\graphicspath{{logos/}}

\usepackage{latex/atlasbiblatex}

\usepackage{epsfig}

\def\ptjetl{p_{\mathrm{T}}^{\text{jet-lead}}}
\def\etg{E_{\mathrm{T}}^{\gamma}}
\def\rapjet{y^{\mathrm{jet}}}
\def\etag{\eta^{\gamma}}
\def\mgj{m^{\gamma-{\mathrm{jet}}}}
\def\ctgj{\cos\theta^*}
\def\dr{\Delta R^{\gamma-{\mathrm{jet}}}}
\def\deltaphigj{\Delta\phi^{\gamma-{\mathrm{jet}}}}
\def\pt{p_{\mathrm{T}}}
\def\pb1{pb$^{-1}$}
\def\fb1{fb$^{-1}$}
\def\qq{q\bar q}
\def\dgen{\frac{{\mathrm{d}}\sigma}{{\mathrm{d}}O}}
\def\dgenn{{\mathrm{d}}\sigma/{\mathrm{d}}O}
\def\dsetg{{\mathrm{d}}\sigma/{\mathrm{d}}\etg}
\def\dspt{{\mathrm{d}}\sigma/{\mathrm{d}}\ptjetl}
\def\dsphi{{\mathrm{d}}\sigma/{\mathrm{d}}\deltaphigj}
\def\dsmgj{{\mathrm{d}}\sigma/{\mathrm{d}}\mgj}
\def\dsctgj{{\mathrm{d}}\sigma/{\mathrm{d}}|\ctgj|}
\def\rbg{R^{\mathrm{bg}}}
\def\ppgj{pp\rightarrow\gamma + {\mathrm{jet}} + X}

\def\mz{m_Z}
\def\as{\alpha_{\mathrm{s}}}
\def\asz{\alpha_{\mathrm{s}}(\mz)}
\def\muR{\mu_{\mathrm{R}}}
\def\muF{\mu_{\mathrm{F}}}
\def\muf{\mu_{\mathrm{f}}}

\def\etiso{E_{\mathrm{T}}^{\mathrm{iso}}}
\def\etisoc{E_{\mathrm{T,cut}}^{\mathrm{iso}}}
\def\etisocut{$4.2\cdot 10^{-3}\cdot \etg +4.8\ {\mathrm{\GeV}}$}
\def\etisocutp{$4.2\cdot 10^{-3}\cdot \etg +10\ {\mathrm{\GeV}}$}
\def\ptjet{p_{\mathrm{T}}^{\text{jet}}}
\def\rapjetl{y^{\text{jet-lead}}}
\def\sher{{\textsc{Sherpa}}}
\def\pyt{{\textsc{Pythia}}}
\def\jetp{{\textsc{Jetphox}}}
\mathchardef\mhyphen="2D

\AtlasTitle{Measurement of the cross section for isolated-photon plus
jet production in {\boldmath $pp$} collisions at {\boldmath $\sqrt
s=13$}~\TeV\ using the ATLAS detector}

\author{The ATLAS Collaboration}

\AtlasJournalRef{Phys. Lett. B 780 (2018) 578}
\AtlasDOI{DOI:10.1016/j.physletb.2018.03.035}
\PreprintIdNumber{CERN-EP-2017-265} 

\AtlasAbstract{
The dynamics of isolated-photon production in association with a jet
in proton--proton collisions at a centre-of-mass energy of $13$~\TeV\
are studied with the ATLAS detector at the LHC using a dataset with an
integrated luminosity of $3.2$~fb$^{-1}$. Photons are required to have
transverse energies above $125$~\GeV. Jets are identified using the
anti-$k_t$ algorithm with radius parameter $R=0.4$ and required to
have transverse momenta above $100$~\GeV.  Measurements of
isolated-photon plus jet cross sections are presented as functions of
the leading-photon transverse energy, the leading-jet transverse
momentum, the azimuthal angular separation between the photon and the
jet, the photon--jet invariant mass and the scattering angle in the
photon--jet centre-of-mass system. Tree-level plus parton-shower
predictions from {\textsc{Sherpa}} and {\textsc{Pythia}} as well as
next-to-leading-order QCD predictions from {\textsc{Jetphox}} and
{\textsc{Sherpa}} are compared to the measurements.
}

\begin{document}

\maketitle

\section{Introduction}
\label{intro}

The production of prompt photons in association with at least one jet
in proton--proton ($pp$) collisions provides a testing ground for
perturbative QCD (pQCD). In $pp$ collisions, all photons that are not
secondaries from hadron decays are considered to be ``prompt''. The
measurements of angular correlations between the photon and the jet
can be used to probe the dynamics of the hard-scattering process. The
dominant source in $pp$ collisions at the LHC is the $qg\rightarrow
q\gamma$ process. These measurements are also useful for tuning Monte
Carlo (MC) models and testing $t$-channel quark
exchange~\cite{np:b875:483,np:b918:257}. Furthermore, precise
measurements of these processes validate the generators used for
background studies in searches for physics beyond the Standard Model
which involve photons, such as the search for new phenomena in final
states with a photon and a  jet~\cite{jhep:1603:041,1709.10440}.

The production of $\ppgj$ events proceeds via two processes: direct,
in which the photon originates from the hard process, and
fragmentation, in which the photon arises from the fragmentation of a
coloured high transverse momentum\footnote{\AtlasCoordFootnote}
($\pt$) parton~\cite{pr:d76:034003,pr:d79:114024}. The direct and
fragmentation contributions are only well defined at leading order
(LO) in QCD; at higher orders this distinction is no longer
possible. These two processes exhibit distinct behaviours in the
observables considered here. Precise measurements test the interplay
of direct and fragmentation processes.

Measurements of prompt-photon production in a final state with
accompanying hadrons necessitate an isolation requirement on the
photon to avoid the large contribution from neutral-hadron decays into
photons. The production of isolated photons in association with jets
in $pp$ collisions at $\sqrt s=7$ and $8$~\TeV\ was studied by the
ATLAS~\cite{pr:d85:092014,np:b875:483,np:b918:257} and
CMS~\cite{pr:d88:112009,jhep:1406:009,jhep:1510:128}
Collaborations. The increase in the centre-of-mass energy of $pp$
collisions at the LHC to $13$~\TeV\ allows the exploration of the
dynamics of photon+jet production in a new regime with the goal of
testing the pQCD predictions at higher energy transfers than achieved
before. It is also possible to investigate whether the data in the new
energy regime are well described by the predictions of parton-shower
event generators, such as \sher~\cite{jhep:0902:007} and
\pyt~\cite{cpc:178:852}.

The dynamics of the underlying processes in $2\rightarrow 2$ hard
collinear scattering can be investigated using the variable
$\theta^*$, where $\ctgj\equiv\tanh(\Delta y/2)$ and $\Delta y$ is the
difference between the rapidities of the two final-state
particles. The variable $\theta^*$ coincides with the scattering polar
angle in the centre-of-mass frame for collinear scattering of massless
particles, and its distribution is sensitive to the spin of the
exchanged particle. For processes dominated by $t$-channel gluon
exchange, such as dijet production in $pp$ collisions, the
differential cross section behaves as $(1-|\ctgj|)^{-2}$ when
$|\ctgj|\rightarrow 1$. In contrast, processes dominated by
$t$-channel quark exchange are expected to exhibit a
$(1-|\ctgj|)^{-1}$ behaviour when $|\ctgj|\rightarrow 1$. This
fundamental prediction of QCD can be tested in photon plus jet
production in $pp$ collisions. The direct-photon contribution,
dominated by $t$-channel quark exchange, is expected to exhibit a
$(1-|\ctgj|)^{-1}$ dependence when $|\ctgj|\rightarrow 1$, whereas
that of fragmentation processes is predicted to be the same as in
dijet production, namely $(1-|\ctgj|)^{-2}$. For both processes, there
are also $s$-channel contributions which are, however, non-singular
when $|\ctgj|\rightarrow 1$. As a result, a measurement of the cross
section for prompt-photon plus jet production as a function of
$|\ctgj|$ provides a handle on the relative contributions of the
direct and fragmentation components as well as the possibility of
testing the dominance of $t$-channel quark exchange.

The results presented here include a study of the kinematics of the
photon plus one-jet system via measurements of the cross sections as
functions of the leading-photon transverse energy ($\etg$) and the
leading-jet transverse momentum ($\ptjetl$). The dynamics of the
photon plus one-jet system are studied by measuring the azimuthal
angular separation between the leading photon and the leading jet
($\deltaphigj$), the invariant mass of the leading photon and the
leading jet ($\mgj$) and $\ctgj$. The distribution in $\mgj$ is
predicted to be monotonically decreasing in QCD due to the absence of
resonances that decay into a photon and a jet. The analysis is
performed using $3.2$~\fb1\ of $\sqrt s=13$~\TeV\ $pp$ collision data
recorded by ATLAS. Next-to-leading-order (NLO) QCD predictions from
\jetp~\cite{jhep:0205:028,pr:d73:094007} and \sher\ as well as the
tree-level predictions of \pyt\ and \sher\ are compared to the
measurements.

\section{ATLAS detector}
\label{detector}
The ATLAS detector~\cite{PERF-2007-01} is a multi-purpose detector
with a forward-backward symmetric cylindrical geometry. It consists of
an inner tracking detector surrounded by a thin superconducting
solenoid, electromagnetic and hadronic calorimeters, and a muon
spectrometer incorporating three large superconducting toroid
magnets. The inner-detector system is immersed in a $2$~T axial
magnetic field and provides charged-particle tracking in the range
$|\eta|<2.5$. The high-granularity silicon pixel detector is closest
to the interaction region and provides four measurements per track;
the innermost layer, known as the insertable
B-layer~\cite{ATLAS-TDR-19}, provides high-resolution hits at small
radius to improve the tracking performance. The pixel detector is
followed by the silicon microstrip tracker, which typically provides
four three-dimensional space point measurements per track. These
silicon detectors are complemented by the transition radiation
tracker, which enables radially extended track reconstruction up to
$|\eta|=2.0$. The calorimeter system covers the range
$|\eta|<4.9$. Within the region $|\eta|<3.2$, electromagnetic (EM)
calorimetry is provided by barrel and endcap high-granularity
lead/liquid-argon (LAr) EM calorimeters, with an additional thin LAr
presampler covering $|\eta|<1.8$ to correct for energy loss in
material upstream of the calorimeters; for $|\eta|<2.5$ the EM
calorimeter is divided into three layers in depth. Hadronic
calorimetry is provided by a steel/scintillator-tile calorimeter,
segmented into three barrel structures within $|\eta|<1.7$, and two
copper/LAr hadronic endcap calorimeters, which cover the region
$1.5<|\eta|<3.2$. The solid-angle coverage is completed out to
$|\eta|=4.9$ with forward copper/LAr and tungsten/LAr calorimeter
modules, which are optimised for EM and hadronic measurements,
respectively. Events are selected using a first-level trigger
implemented in custom electronics, which reduces the maximum event
rate of 40~MHz to a design value of 100~kHz using a subset of detector
information. Software algorithms with access to the full detector
information are then used in the high-level trigger to yield a
recorded event rate of about 1~kHz~\cite{epj:c77:317}.

\section{Data sample and Monte Carlo simulations}
\label{datmc}
The data used in this analysis were collected with the ATLAS detector
during the $pp$ collision running period of 2015, when the LHC
operated with a bunch spacing of $25$~ns and at a centre-of-mass
energy of $13$~\TeV. Only events taken during stable beam conditions
and satisfying detector and data-quality requirements, which include
the calorimeters and inner tracking detectors being in nominal
operation, are considered. The average number of $pp$ interactions per
bunch crossing in the dataset is $13$.  The total integrated
luminosity of the collected sample is {\mbox{$3.16 \pm 0.07$~\fb1}}.
The uncertainty in the integrated luminosity is $2.1\%$ and is
derived, following a methodology similar to that detailed in
Ref.~\cite{epj:c76:653}, from a calibration of the luminosity scale
using $x$--$y$ beam-separation scans performed in August 2015.

Samples of MC events were generated to study the characteristics of
signal events. The MC programs \pyt~8.186 and \sher~2.1.1 were used to
generate the simulated events. In both generators, the partonic
processes were simulated using tree-level matrix elements, with the
inclusion of initial- and final-state parton showers. Fragmentation
into hadrons was performed using the Lund string
model~\cite{prep:97:31} in the case of  \pyt, and in \sher\ by a
modified version of the cluster model~\cite{epj:c36:381}. The LO
NNPDF2.3~\cite{np:b867:244} parton distribution functions (PDF) set
was  used for \pyt\ while the NLO CT10~\cite{pr:d82:074024} PDF set
was used for \sher\ to parameterise the proton structure. Both samples
include a simulation of the underlying event (UE). The event-generator
parameters were set according to the ATLAS  2014 tune series (A14
tune) for \pyt~\cite{ATL-PHYS-PUB-2014-021} and to the tune developed
in conjunction with the NLO CT10 PDF set for \sher. The \pyt\
simulation of the signal includes LO photon-plus-jet events from both
direct processes (the hard subprocesses  $qg\rightarrow q\gamma$ and
$\qq\rightarrow g\gamma$, called the ``hard'' component) and photon
bremsstrahlung in LO QCD dijet events (called the ``bremsstrahlung''
component). The bremsstrahlung component is modelled by final-state
QED radiation arising from calculations of all $2\rightarrow 2$ QCD
processes. In the particle-level phase space of the presented
measurements the fraction of the bremsstrahlung component decreases
from $35\%$ at $\etg=125$~\GeV\ to $15\%$ at $\etg=1$~\TeV. The \sher\
samples were generated with LO matrix elements for photon-plus-jet
final states with up to three additional partons ($2\rightarrow n$
processes with $n$ from $2$ to $5$); the matrix elements were merged
with the \sher\ parton shower using the ME+PS@LO
prescription~\cite{jhep:0905:053}. The bremsstrahlung component is
accounted for in \sher\ through the matrix elements of $2\rightarrow
n$ processes with $n\geq 3$. In the generation of the \sher\ samples,
a requirement on the photon isolation at the matrix-element level was
imposed using the criterion defined in Ref.~\cite{pl:b429:369}. This
criterion, commonly called Frixione's criterion, requires the total
transverse energy inside a cone of size ${\cal V}$ around the
generated final-state photon, excluding the photon itself, to be below
a certain threshold,
$E_{\mathrm{T}}^{\mathrm{max}}({\cal V})=\epsilon\etg ((1-\cos{{\cal V}})/(1-\cos{{\cal R}}))^{n}$,
for all ${\cal V} < {\cal R}$. The parameters for the threshold were
chosen to be ${\cal R}=0.3$, $n=2$ and $\epsilon=0.025$. The \sher\
predictions from this computation are referred to as LO \sher. 

All the samples of generated events were passed through the
{\textsc{Geant4}}-based~\cite{nim:a506:250} ATLAS detector and trigger
full simulation programs~\cite{epj:c70:823}. They are reconstructed
and analysed with the same program chain as the data. Pile-up from
additional $pp$ collisions  in the same and neighbouring bunch
crossings was simulated by overlaying each MC event with a variable
number of simulated inelastic $pp$ collisions generated using
\pyt~8.153 with the ATLAS set of tuned parameters for minimum bias
events (A2 tune)~\cite{ATL-PHYS-PUB-2012-003}. The MC events are
weighted to reproduce the distribution of the average number of
interactions per bunch crossing ($\langle\mu\rangle$) observed in the
data, referred to as ``pile-up reweighting''. In this procedure, the
$\langle\mu\rangle$ value from the data is divided by a factor of
$1.16\pm 0.07$, a rescaling which makes the number  of reconstructed
primary vertices agree better between data and simulation and
reproduces the visible cross section of inelastic $pp$ collisions as
measured in the data~\cite{nc:2:463}.

\section{Event selection}
\label{evesel}
Events were recorded using a single-photon trigger, with a transverse
energy threshold of $120$~\GeV\ and ``loose'' identification
requirements based on the shower shapes in the second layer of the EM
calorimeter as well as on  the energy leaking into the hadronic
calorimeter from the EM calorimeter~\cite{epj:c77:317}. Events are
required to have a reconstructed primary vertex. If multiple primary
vertices are reconstructed, the one with the highest sum of the
$\pt^2$ of the associated tracks is selected as the primary vertex.

\nopagebreak
Photon candidates are reconstructed from clusters of energy deposited
in the EM calorimeter and  classified~\cite{epj:c76:666} as
unconverted photons (candidates without a matching track or matching
reconstructed conversion vertex in the inner detector) or converted
photons (candidates with a matching reconstructed conversion vertex or
a matching track consistent with originating from a photon
conversion). The measurement of the photon energy is based on the
energy collected in calorimeter cells in an area of size
$\Delta\eta\times\Delta\phi=0.075\times 0.175$ in the barrel and
$0.125\times 0.125$ in the endcaps. A dedicated energy
calibration~\cite{epj:c74:3071} is then applied to the candidates to
account for upstream energy loss and both lateral and longitudinal
leakage. The photon identification is based primarily on shower shapes
in the calorimeter~\cite{epj:c76:666}. An initial selection is derived
using the information from the hadronic calorimeter and the lateral
shower shape in the second layer of the EM calorimeter, where most of
the photon energy is contained. The final tight selection applies
stringent criteria~\cite{epj:c76:666} to the same variables used in
the initial selection, separately for converted and unconverted photon
candidates. It also places requirements on the shower shape in the
finely segmented first calorimeter layer to ensure the compatibility
of the measured shower profile with that originating from a single
photon impacting the calorimeter. When applying the photon
identification criteria to simulated events, corrections are made for
small differences in the average values of the shower-shape variables
between data and simulation. Events with at least one photon candidate
with calibrated $\etg>125$~\GeV, where the trigger is maximally
efficient, and $|\etag|<2.37$ are selected. Candidates in the region
$1.37<|\etag|<1.56$, which includes the transition region between the
barrel and endcap calorimeters, are not considered. The photon
candidate is required to be isolated based on the amount of transverse
energy inside a cone of size $\Delta R=0.4$ in the $\eta$--$\phi$
plane around the photon candidate, excluding an area of size
$\Delta\eta \times \Delta\phi = 0.125 \times 0.175$ centred on the
photon. The isolation transverse energy is computed from topological
clusters of calorimeter cells~\cite{epj:c77:490} and is denoted by
$\etiso$. Topological clusters are built from neighbouring calorimeter
cells containing energy significantly above a noise threshold that is
estimated from measurements of calorimeter electronic noise and
simulated pile-up noise. The measured value of $\etiso$ is corrected
for the expected leakage of the photon energy into the isolation cone
as well as for the estimated contributions from the UE and
pile-up~\cite{pr:d83:052005,pl:b706:150}. The corrections for pile-up
and the UE are computed simultaneously on an event-by-event basis
using the jet-area method~\cite{jhep:0804:005,jhep:1004:065} as
follows: the $k_\perp$ jet algorithm~\cite{pr:d48:3160,np:b406:187}
with jet radius $R=0.5$ is used to reconstruct all jets, taking
topological clusters of calorimeter cells as input; no explicit
transverse momentum threshold is applied. The ambient transverse
energy density for the event ($\rho$), from pile-up and the UE, is
computed using the median of the distribution of the ratio of the
jet's transverse energy to its area. Finally, $\rho$ is multiplied by
the area of the isolation cone to compute the correction to
$\etiso$. The combined correction is typically $2$~\GeV\ and depends
weakly on $\etg$. In addition, for simulated events, data-driven
corrections to $\etiso$ are applied such that the peak position in the
$\etiso$ distribution coincides in data and simulation. After all
these corrections, $\etiso$ is required to be less than
$\etisoc\equiv$~\etisocut~\cite{jhep:1608:005}. The isolation
requirement significantly reduces the main background, which consists
of multi-jet events where one jet typically contains a $\pi^0$ or
$\eta$ meson that carries most of the jet energy and is misidentified
as a photon because it decays into an almost collinear photon pair. A
small fraction of the events contain more than one photon candidate
satisfying the selection criteria. In such events, the highest-$\etg$
(leading) photon is considered for further study.

Jets are reconstructed using the anti-$k_t$
algorithm~\cite{jhep:0804:063,epj:c72:1896} with a radius parameter
$R=0.4$, using topological clusters as input. The calorimeter cell
energies are measured at the EM scale, corresponding to the energy
deposited by electromagnetically interacting particles. The jet
four-momenta are computed from the sum of the jet-constituent
four-momenta, treating each as a four-vector with zero mass. The jets
are then further calibrated using the method described in
Ref.~\cite{pr:d96:072002} and these jets are referred to as
detector-level jets. The four-momentum of each jet is recalculated to
point to the selected primary vertex of the event rather than the
centre of the detector. The contribution from the UE and pile-up is
then subtracted on a jet-by-jet basis using the jet-area method. A
jet-energy calibration is derived from MC simulations as a correction
relating the calorimeter response to the true jet energy. To determine
these corrections, the jet reconstruction procedure applied to the
topological clusters is also applied to the generated stable
particles, which are defined as those with a decay length of
$c\tau>10$~mm, excluding muons and neutrinos; these jets are referred
to as particle-level jets. In addition, sequential jet corrections,
derived from MC simulated events and using global properties of the
jet such as tracking information, calorimeter energy deposits and muon
spectrometer information, are
applied~\cite{ATLAS-CONF-2015-002}. Finally, the detector-level jets
are further calibrated with additional correction factors derived in
situ from a combination of $\gamma+{\mathrm{jet}}$, $Z+{\mathrm{jet}}$
and multi-jet $p_{\mathrm{T}}$ balance
methods~\cite{epj:c75:17,ATLAS-CONF-2015-037}.\footnote{The effect of
  the correlation between the events used in the in situ $\gamma+$jet
  analysis and the events selected here has a negligible effect on the
  experimental uncertainties associated to the measurements.}

Jets reconstructed from calorimeter signals not originating from a
$pp$ collision are rejected by applying jet-quality
criteria~\cite{epj:c75:17,epj:c73:2304}. These criteria suppress
spurious jets from electronic noise in the calorimeter, cosmic rays
and beam-related backgrounds. Remaining jets are required to have
calibrated transverse momenta greater than $60$~\GeV\ and rapidity
$|\rapjet|<2.37$. Jets overlapping with the candidate photon are not
considered if the jet axis lies within a cone of size $\Delta R=0.8$
around the photon candidate; this requirement prevents any overlap
between the photon isolation cone ($\Delta R=0.4$) and the jet cone
($\Delta R=0.4$). Finally, the event is retained if the jet with
highest transverse energy (leading jet) has $\ptjet>100$~\GeV.

The total number of data events selected by using the requirements
discussed above is $895\ 726$.  This sample of events is used to
measure the cross section as a function of $\etg$, $\ptjetl$ and
$\deltaphigj$. For the measurements of the cross sections as functions
of $\mgj$ and $|\ctgj|$, the additional constraints
$|\etag+\rapjetl|<2.37$, $|\ctgj|<0.83$ and $\mgj>450$~\GeV\ are
imposed to remove the bias~\cite{np:b875:483,np:b918:257} due to the
rapidity and transverse-momentum requirements on the photon and the
jet;\footnote{The first two constraints avoid the bias induced by
  requirements on $\etag$ and $\rapjetl$, yielding slices of $\ctgj$
  with the same length along the $\etag+\rapjetl$ axis. The third
  constraint avoids the bias due to the $\etg>125$~\GeV\ requirement.}
the number of events selected in the data after these additional
requirements is $137\ 738$.

\section{Background estimation and signal extraction}
\label{bacsub}
After the requirements on photon identification and isolation are
applied to the sample of events, there is still a residual background
contribution, which arises primarily from jets identified as photons
in multi-jet events. This background contribution is estimated and
subtracted bin-by-bin using a data-driven technique which makes use of
the same two-dimensional sideband method employed in a previous
analysis~\cite{pl:b770:473}. In this approach, the photon is
classified as:
\begin{itemize}
\item ``Isolated'', if $\etiso<\etisoc$.
\item ``Non-isolated'', if $\etiso>\etisoc +2\ {\mathrm{\GeV}}$ and
  $\etiso< 50$~\GeV. The non-isolated region is separated by $2$~\GeV\
  from the isolated region to reduce the signal contamination. The
  upper bound is applied to avoid highly non-isolated
  photons.\footnote{In this way, the determination of the signal yield
    does not depend on the description by the MC generators of the
    distribution of $\etiso$ for prompt photons with high values of
    $\etiso$.} 
\item ``Tight'', if it satisfies the tight photon identification criteria.
\item ``Non-tight'', if it fails at least one of four tight
  requirements on the shower-shape variables computed from the energy
  deposits in the first layer of the EM calorimeter, but satisfies the
  tight requirement on the total lateral shower width in the first
  layer and all the other tight identification criteria in other
  layers~\cite{epj:c76:666}.
\end{itemize}

\begin{figure}[h]
\setlength{\unitlength}{1.0cm}
\begin{picture} (18.0,6.0)
\put (0.0,-0.4){\includegraphics[width=6.8cm,height=6.8cm]{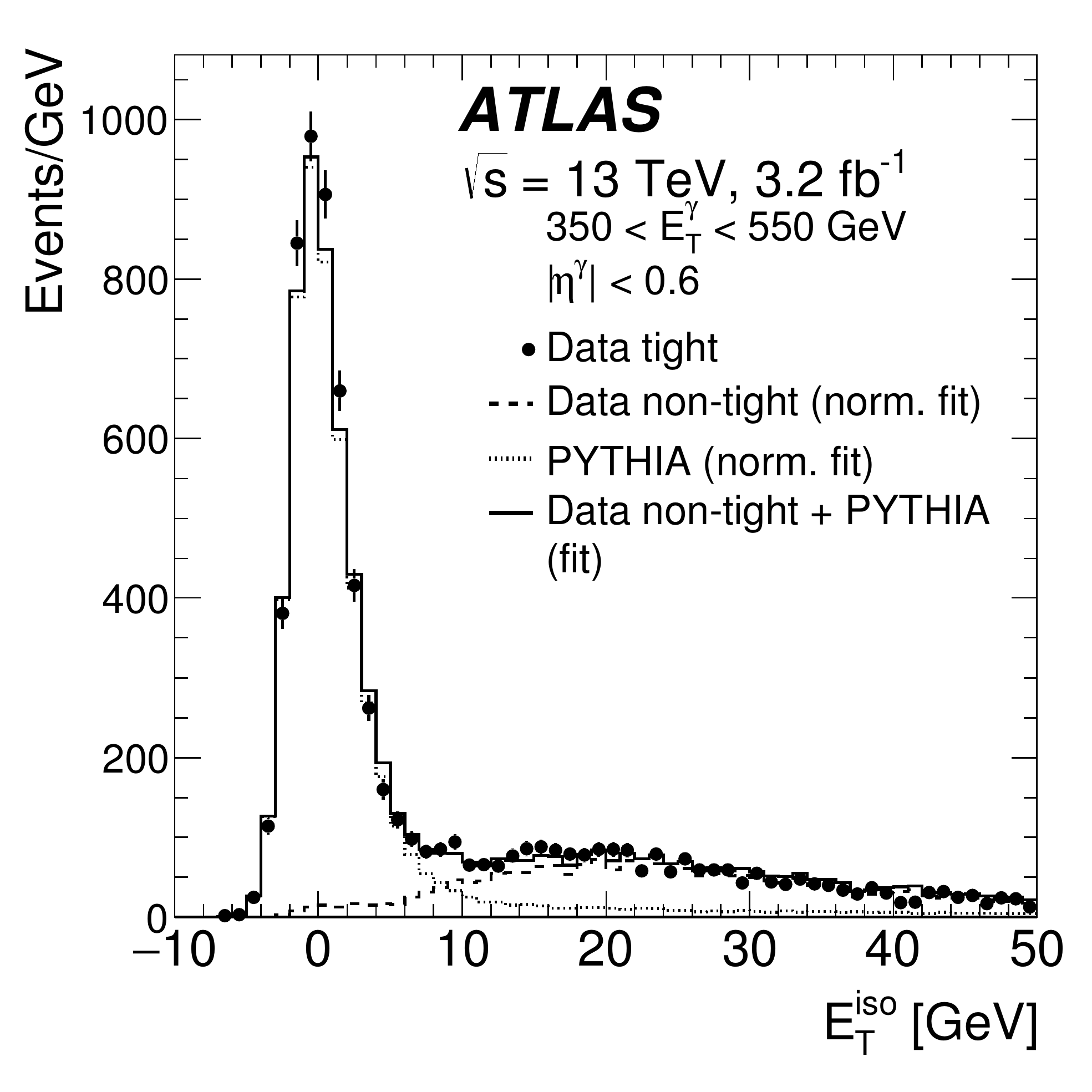}}
\put (8.0,-0.4){\includegraphics[width=6.8cm,height=6.8cm]{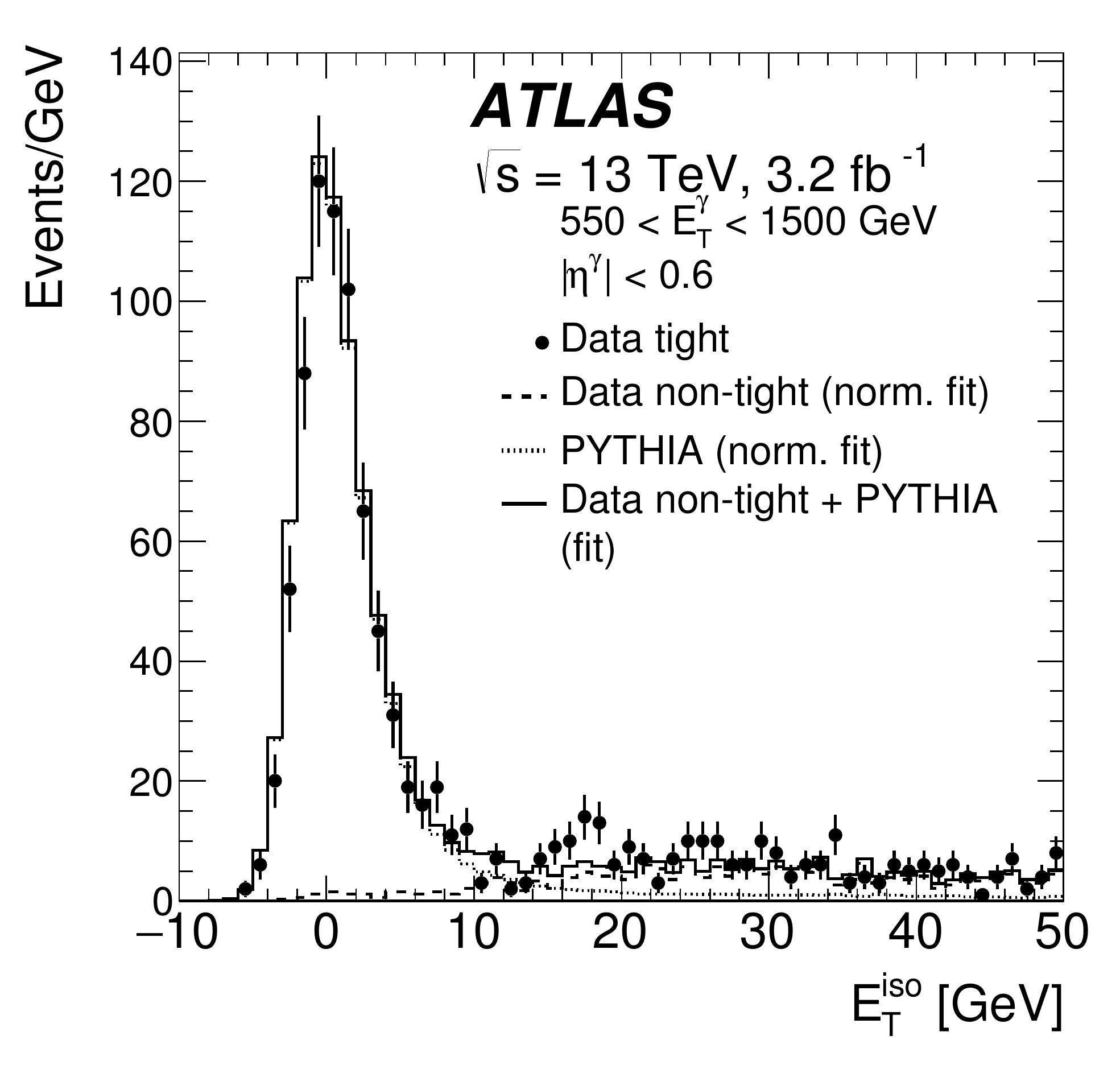}}
\end{picture}
\caption
{
  $\etiso$ distribution for tight (black dots) and non-tight (dashed
  histogram, normalised according to the fit, see text) photon
  candidates in data with $|\etag|<0.6$ in different $\etg$
  regions. The MC simulation of the signal using \pyt\ is also shown
  (dotted histogram). The solid histogram is the sum of the
  contributions of the MC simulation of the signal using \pyt\  and
  that of the non-tight photon candidates normalised according to the
  fit.
}
\label{fig0}
\end{figure}

The distributions in $\etiso$ for tight and non-tight photon
candidates with $|\eta^{\gamma}|<0.6$ in the data are shown separately
in Figure~\ref{fig0} for two regions in $\etg$. The MC simulation of
the prompt-photon signal using \pyt\ is also shown. A fit of the sum
of the distributions of the \pyt\ signal photons and the non-tight
photon candidates to the distribution of the tight photon candidates
is also included. A clear signal of prompt photons centred at $\etiso$
about zero is observed.

For the estimation of the background contamination in the signal
region a two-dimensional plane is formed by $\etiso$ and a binary
variable (``tight'' vs. ``non-tight'' photon candidate) since these
two variables are expected to be largely uncorrelated for background
events. In this plane, four regions are defined: the ``signal'' region
($A$), containing tight and isolated photon candidates; the
``non-isolated'' background control region ($B$), containing tight and
non-isolated photon candidates; the ``non-tight'' background control
region ($C$), containing isolated and non-tight photon candidates; the
background control region containing non-isolated and non-tight photon
candidates ($D$).

The signal yield $N_A^{\mathrm{sig}}$ in region $A$ is estimated by
using the relation
\begin{equation}
  N_A^{\mathrm{sig}}=N_A-R^{\mathrm{bg}}\cdot(N_B-f_B N_A^{\mathrm{sig}})\cdot
  \frac{(N_C-f_C N_A^{\mathrm{sig}})}{(N_D-f_D N_A^{\mathrm{sig}})},
  \label{eqone}
  \end{equation}
where $N_K$, with $K=A,B,C,D$, is the number of events in region $K$
and $R^{\mathrm{bg}}=N_A^{\mathrm{bg}}\cdot N_D^{\mathrm{bg}}/(N_B^{\mathrm{bg}}\cdot N_C^{\mathrm{bg}})$
is the so-called background correlation and is taken as
$R^{\mathrm{bg}}=1$ for the nominal results; $N_K^{\mathrm{bg}}$ with
$K=A,B,C,D$ is the unknown number of background events in each
region. Equation~(\ref{eqone}) takes into account the expected number
of signal events in each of the three background control regions
($N_K^{\mathrm{sig}}$) via the signal leakage fractions,
$f_K=N_K^{\mathrm{sig}}/N_A^{\mathrm{sig}}$ with $K=B,C,D$, which are
estimated using the MC simulations of the signal.

The only assumption underlying Eq.~(\ref{eqone}) is that the isolation
and identification variables are uncorrelated for background events,
thus $R^{\mathrm{bg}}=1$. This assumption is verified in data
typically within $\pm 10\%$ in validation regions,\footnote{The
  validation regions are defined within the control regions $B$ and
  $D$ by splitting each of them into two subregions.} which are
dominated by background. Deviations of $R^{\mathrm{bg}}$ from unity in
the validation regions, after accounting for signal leakage using
either the \pyt\ or LO \sher\ simulations, are propagated  through
Eq.~(\ref{eqone}) and taken as systematic uncertainties. The signal
purity, defined as $N_A^{\mathrm{sig}}/N_A$, is above $90\%$ in all
bins of the measured distributions. The use of \pyt\ or LO \sher\ to
extract the signal leakage fractions lead to similar signal purities;
the difference in the signal purity is taken as a systematic
uncertainty.

The background from electrons misidentified as photons, mainly
produced in Drell--Yan $Z^{(*)}/\gamma^*\rightarrow e^+e^-$ and
$W^{(*)}\rightarrow e\nu$ processes, is also studied. Such
misidentified electrons are largely suppressed by the photon
selection. This background is estimated using MC samples of fully
simulated events and found to be negligible in the phase-space region
of the analysis presented here. These processes were simulated using
the \sher~2.2.1 generator. Matrix-elements were calculated for up to
two additional partons at NLO and up to four partons at
LO~\cite{jhep:0202:044,prl:108:111601}.

\section{Fiducial phase space and unfolding}

\subsection{Fiducial phase space}
\label{fiducial}
The cross sections are unfolded to a phase-space region close to the
applied event selection. The fiducial phase-space region is defined at
the particle level. A summary of the requirements at particle level
that define the fiducial phase-space region of the measurements is
given in Table~\ref{table1}. The cross sections as functions of
$|\ctgj|$ and $\mgj$ are measured in a fiducial phase-space region
with additional requirements, as detailed in the last row of
Table~\ref{table1}. The particle-level isolation requirement on the
photon is built by summing the transverse energy of all stable
particles (see Section~\ref{evesel}), except for muons and neutrinos,
in a cone of size $\Delta R=0.4$ around the photon direction after the
contribution from the UE is subtracted. The same underlying-event
subtraction procedure used at the reconstruction level is applied at
the particle level. The particle-level requirement on $\etiso$ is
optimised to best match the acceptance at reconstruction level using
the \pyt\ and LO \sher\ MC samples by comparing the calorimeter
isolation transverse energy with the particle-level isolation
transverse energy on an event-by-event basis. The particle-level
requirement on $\etiso$ thus optimised is
$\etiso({\mathrm{particle}})<$~\etisocutp; the same requirement is
obtained whether \pyt\ or LO \sher\ is used. Particle-level jets are
reconstructed using the anti-$k_t$ jet algorithm with radius parameter
$R=0.4$ and are built from stable particles, excluding muons and
neutrinos. At particle level, the particles associated with the
overlaid $pp$ collisions (pile-up) are not considered. 

\vspace{-0.3cm}
\begin{table}[h]
\caption{
  Summary of the requirements at particle level that define the
  fiducial phase-space region of the measurements.
}
\setlength{\unitlength}{1.0cm}
\begin{picture} (18.0,4.0)
\put (0.0,1.5){\centerline{\scalebox{1.0}{
\begin{tabular}{|l|} \hline
  {\textbf{Requirements on photons}} \\ 
  $\etg>125$~\GeV, $|\etag|<2.37$ (excluding $1.37<|\etag|<1.56$) \\
  $\etiso<$~\etisocutp \\ \hline
  {\textbf{Requirements on jets}} \\ 
  anti-$k_t$ algorithm with $R=0.4$  \\
  the leading jet within $|\rapjet|<2.37$ and $\dr>0.8$ is selected\\
  $\ptjetl>100$~\GeV \\ \hline
  {\textbf{UE subtraction using $k_\perp$ algorithm with $R=0.5$ (cf. Section~\ref{evesel})}} \\ \hline 
  {\textbf{Additional requirements for $\dsmgj$ and $\dsctgj$}} \\ 
  $|\etag+\rapjetl|<2.37$, $|\ctgj|<0.83$ and $\mgj>450$~\GeV\  \\ \hline
\end{tabular}
}}}
\end{picture}
\label{table1}
\end{table}

\subsection{Unfolding}
\label{cor}
The distributions of the background-subtracted signal yield as
functions of $\etg$, $\ptjetl$, $\deltaphigj$, $\mgj$ and $|\ctgj|$
are used to measure the corresponding differential cross sections for
isolated-photon plus jet production. The distributions are unfolded to
the particle level using MC samples of events via a bin-by-bin
technique which corrects for resolution effects and the efficiency of
the photon and jet reconstruction through the formula

\begin{equation}
  \dgen(i)=\frac{N_A^{\mathrm{sig}}(i)\ C^{\mathrm{MC}}(i)}{{\cal L}\ \Delta O(i)}, \nonumber
  \label{eqthree}
\end{equation}
where $\dgenn(i)$ is the cross section as a function of observable $O$
in bin $i$, $N_A^{\mathrm{sig}}(i)$ is the number of
background-subtracted data events in bin $i$, $C^{\mathrm{MC}}(i)$ is
the correction factor in bin $i$, ${\cal L}$ is the integrated
luminosity and $\Delta O(i)$ is the width of bin $i$. The correction
factors are computed from the MC samples as
$C^{\mathrm{MC}}(i)=N^{\mathrm{MC}}_{\mathrm{part}}(i)/N^{\mathrm{MC}}_{\mathrm{reco}}(i)$,
where $N^{\mathrm{MC}}_{\mathrm{part}}(i)$ is the number of events
that satisfy the kinematic constraints of the phase-space region at
the particle level, and $N^{\mathrm{MC}}_{\mathrm{reco}}(i)$ is the
number of events which meet all the selection criteria at the
reconstruction level.

The distributions of the signal yields as functions of $\etg$,
$\ptjetl$, $\deltaphigj$, $\mgj$ and $|\ctgj|$ in data after
background subtraction are well described by the LO \sher\ MC
simulations, but some differences are observed when compared to the
\pyt\ MC simulations, in particular in the tail of the $\ptjetl$
distribution. A better description of the data distributions as
functions of $\etg$, $\ptjetl$, $\deltaphigj$, $\mgj$ and $|\ctgj|$ by
\pyt\ is obtained by increasing/decreasing the relative contribution
from direct processes with respect to bremsstrahlung
processes~\cite{np:b875:483,np:b918:257}; the resulting \pyt\
simulations are referred to as \pyt-adjusted simulations.

The unfolded cross sections are measured using the signal leakage
fractions from \pyt-adjusted simulations since these include an
unbiased sample of non-isolated photons,\footnote{In the LO \sher\
  samples, the application of the Frixione's criterion to the photon
  isolation at matrix-element level prevents the radiated photon from
  being close to a parton. In the \pyt\ samples, the bremsstrahlung
  component is simulated with a parton shower approach and, as a
  result, the radiated photon can be close to a parton.} and the
correction factors, $C^{\mathrm{MC}}$, from LO \sher\ since these
simulations give somewhat better agreement with the data distributions
as functions of $\etg$, $\ptjetl$, $\deltaphigj$, $\mgj$ and
$|\ctgj|$. The correction factors vary between $1.08$ and $1.21$
depending on the observable. The results of the bin-by-bin unfolding
procedure are checked with an iterative Bayesian unfolding
method~\cite{nim:a362:487} based on LO \sher\ simulations, giving
consistent results.

\section{Uncertainties in the cross-section measurements}
\label{syst}
{\textbf{Photon energy scale and resolution.}}
A detailed assessment of the uncertainties in the photon energy scale
and resolution is made using the method reported in
Ref.~\cite{pl:b770:473}. The photon energy scale uncertainties come
mostly from calibration studies using 8~\TeV\
data~\cite{epj:c74:3071}, with additional systematic uncertainties to
take into account the differences between the 2012 and 2015
configurations. The uncertainties are split into independent
components to account for correlations of the uncertainties between
different bins of the measured cross sections. The individual sources
of uncertainty are varied  by $\pm 1\sigma$ in the MC simulations and
propagated through the analysis separately to maintain the full
information about the correlations of the uncertainties between
different bins of the measured cross sections. The impact of the
photon energy resolution uncertainty is much smaller than that of the
photon energy scale uncertainty. The resulting uncertainty in the
measured cross sections is obtained by adding in quadrature all the
individual components and increases from $1\%$ at $\etg=125$~\GeV\ to
$4.5\%$ at $\etg\sim 1.5$~\TeV.

{\textbf{Jet energy scale and resolution.}}
A detailed assessment of the uncertainties in the jet energy scale and
resolution is made using the method reported in
Ref.~\cite{pr:d96:072002}. The individual sources of
uncertainty~\cite{pr:d96:072002} are varied  by $\pm 1\sigma$ in the MC
simulations and propagated through the analysis separately, to
maintain the full information about the correlations of the
uncertainties between different bins of the measured cross
sections. The resulting uncertainty in the measured cross sections is
obtained by adding in quadrature all the individual components and
increases from $1.9\%$ at $\ptjetl=100$~\GeV\ to $7.5\%$ at
$\ptjetl\sim 1$~\TeV.

{\textbf{Parton-shower and hadronisation model dependence.}}
The effects due to the parton-shower and hadronisation models on the
signal purity and detector-to-particle-level correction factors are
studied separately. The effects on the signal purity are estimated as
the differences observed between the nominal results and those
obtained using either the (non-adjusted) \pyt\ or LO \sher\ MC samples
for the determination of the signal leakage fractions. The difference
between the nominal results and those obtained using the \pyt-adjusted
MC samples for the determination of the unfolding correction factors
is taken as a systematic uncertainty. The resulting uncertainties in
the measured cross sections are typically smaller than $2\%$.

{\textbf{Photon identification efficiency.}}
The uncertainty in the photon identification efficiency is estimated
from the effect of differences between shower-shape variable
distributions in data and simulation. From the studies presented in
Refs.~\cite{epj:c76:666,ATL-PHYS-PUB-2016-014}, this procedure is
found to provide a conservative estimate of the uncertainties. The
resulting uncertainty in the measured cross sections is in the range
$1$--$2\%$. The effects on the measured cross sections due to the
uncertainty in the photon reconstruction efficiency, which are
evaluated by repeating the full analysis using a different detector
simulation with increased material in front of the calorimeter, are
found to be negligible.

{\textbf{Photon isolation modelling.}}
The differences between the nominal results and those obtained without
applying the data-driven corrections to $\etiso$ in simulated events
are taken as systematic uncertainties in the measurements due to the
modelling of $\etiso$ in the MC simulation. The resulting uncertainty
in the measured cross sections is less than $1.1\%$.

{\textbf{Definition of the background control regions.}}
The estimation of the background contamination in the signal region is
affected by the choice of background control regions. The control
regions $B$ and $D$ are defined by the lower and upper limits on
$\etiso$ and the choice of inverted photon identification variables
used in the selection of non-tight photons. To study the dependence on
the specific choices, these definitions are varied over a wide
range. The lower limit on $\etiso$ in regions $B$ and $D$ is varied by
$\pm 1$~\GeV, which is larger than any difference between data and
simulations and still provides a sufficient sample to perform the
data-driven subtraction. The upper limit on $\etiso$ in regions $B$
and $D$ is removed. The resulting uncertainty in the measured cross
sections is negligible. Likewise, the choice of inverted photon
identification variables is varied. The analysis is repeated using
different sets of variables: tighter (looser) identification criteria
are defined by applying tight requirements to an extended (restricted)
set of shower-shape variables in the first calorimeter
layer~\cite{epj:c76:666,ATL-PHYS-PUB-2016-014}. The resulting
uncertainty in the measured cross sections is smaller than $1.3\%$.

{\textbf{Photon identification and isolation correlation in the
    background.}}
The photon isolation and identification variables used to define the
plane in the two-dimensional sideband method to subtract the
background are assumed to be independent for background events
($\rbg=1$ in Eq.~(\ref{eqone})). Any correlation between these
variables affects the estimation of the purity of the signal sample
and leads to systematic uncertainties in the background-subtraction
procedure. A range in $\rbg$ is set to cover the deviations from unity
measured in the validation regions after subtracting the signal
leakage with either \pyt-adjusted or LO \sher\ MC samples. The
resulting uncertainty in all measured cross sections is less than
$2\%$.

{\textbf{Pile-up.}}
The uncertainty is estimated by changing the nominal rescaling factor
of $1.16$ to $1.09$ or $1.23$ and re-evaluating the reweighting
factors. The resulting uncertainty in the measured cross sections is
typically less than $0.5\%$.

{\textbf{Unfolding procedure.}}
The uncertainty is estimated by comparing the nominal results with
those obtained by unfolding with LO \sher\ MC samples reweigthed to
match the data distributions. The resulting uncertainty in the
measured cross sections is negligible.

\begin{figure}[h]
\setlength{\unitlength}{1.0cm}
\begin{picture} (18.0,20.0)
\put (0.0,14.0){\includegraphics[width=7cm,height=7cm]{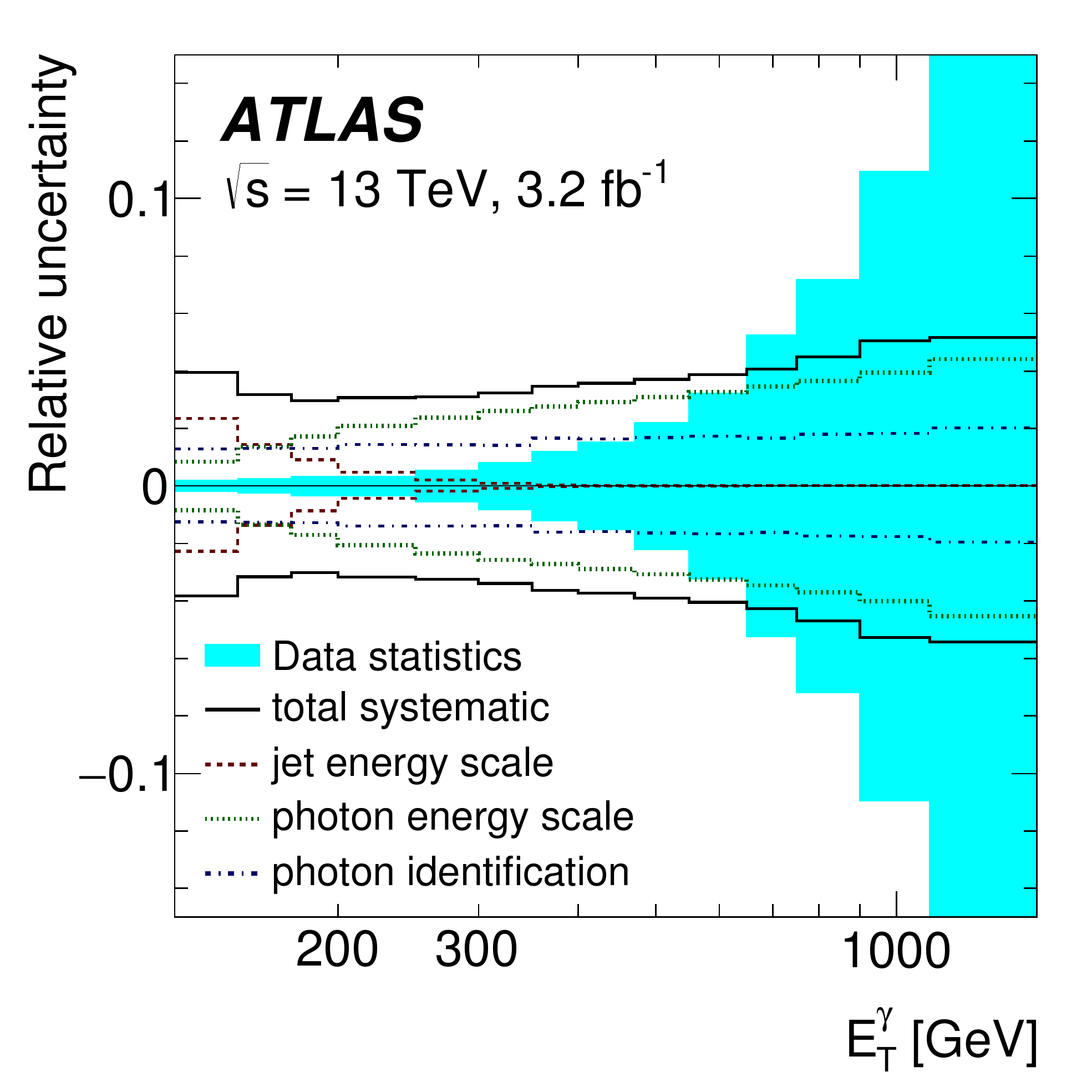}}
\put (8.0,14.0){\includegraphics[width=7cm,height=7cm]{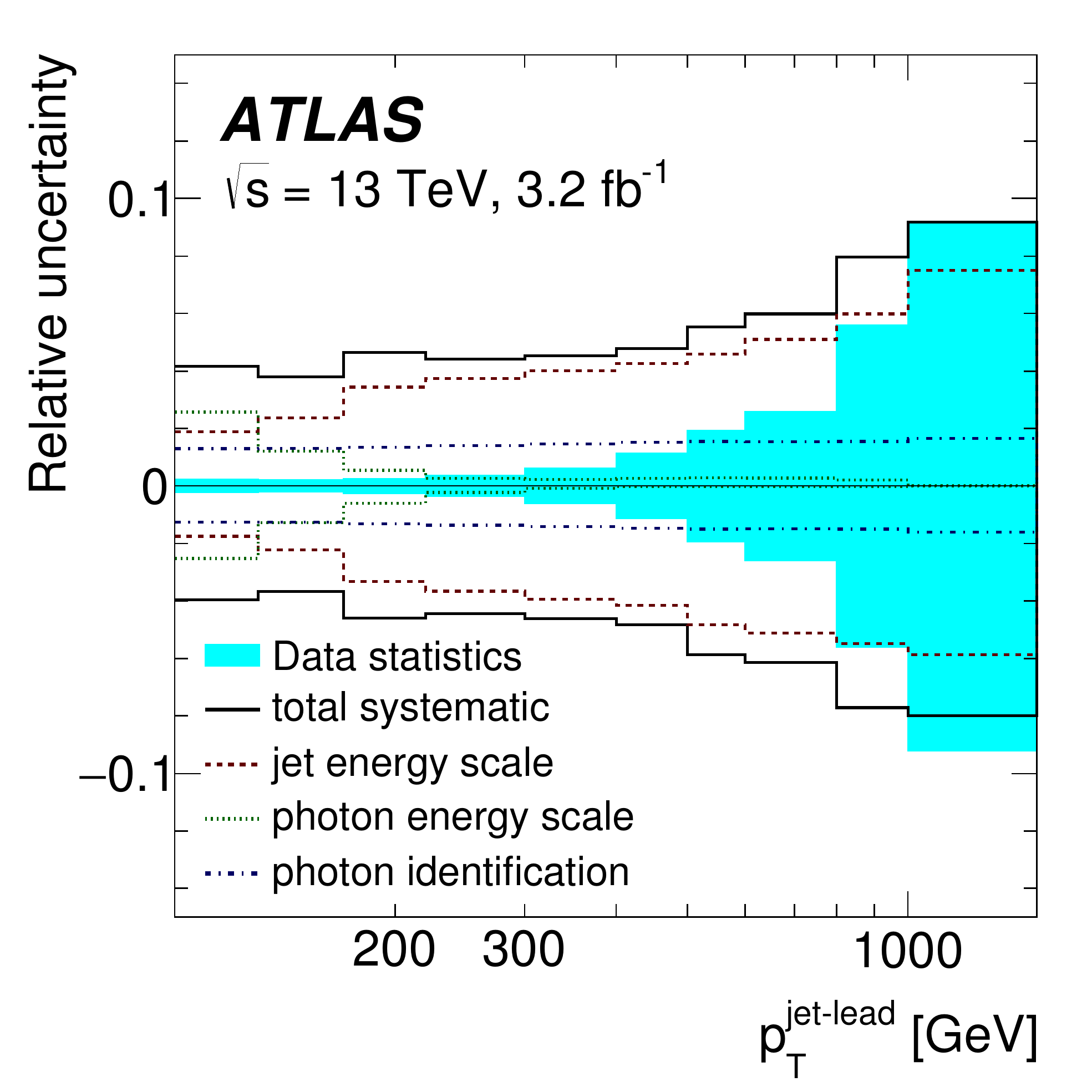}}
\put (0.0,7.0){\includegraphics[width=7cm,height=7cm]{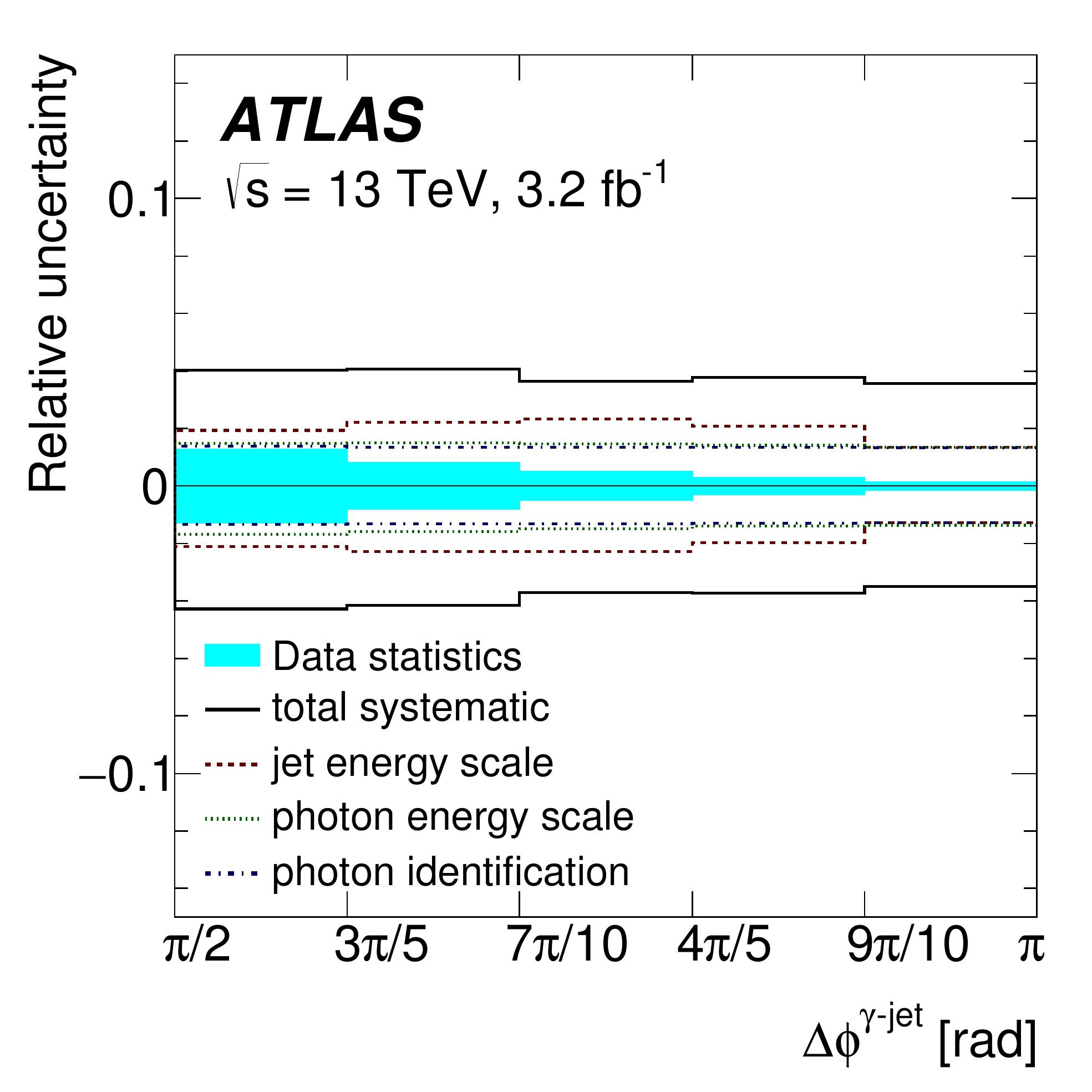}}
\put (0.0,0.0){\includegraphics[width=7cm,height=7cm]{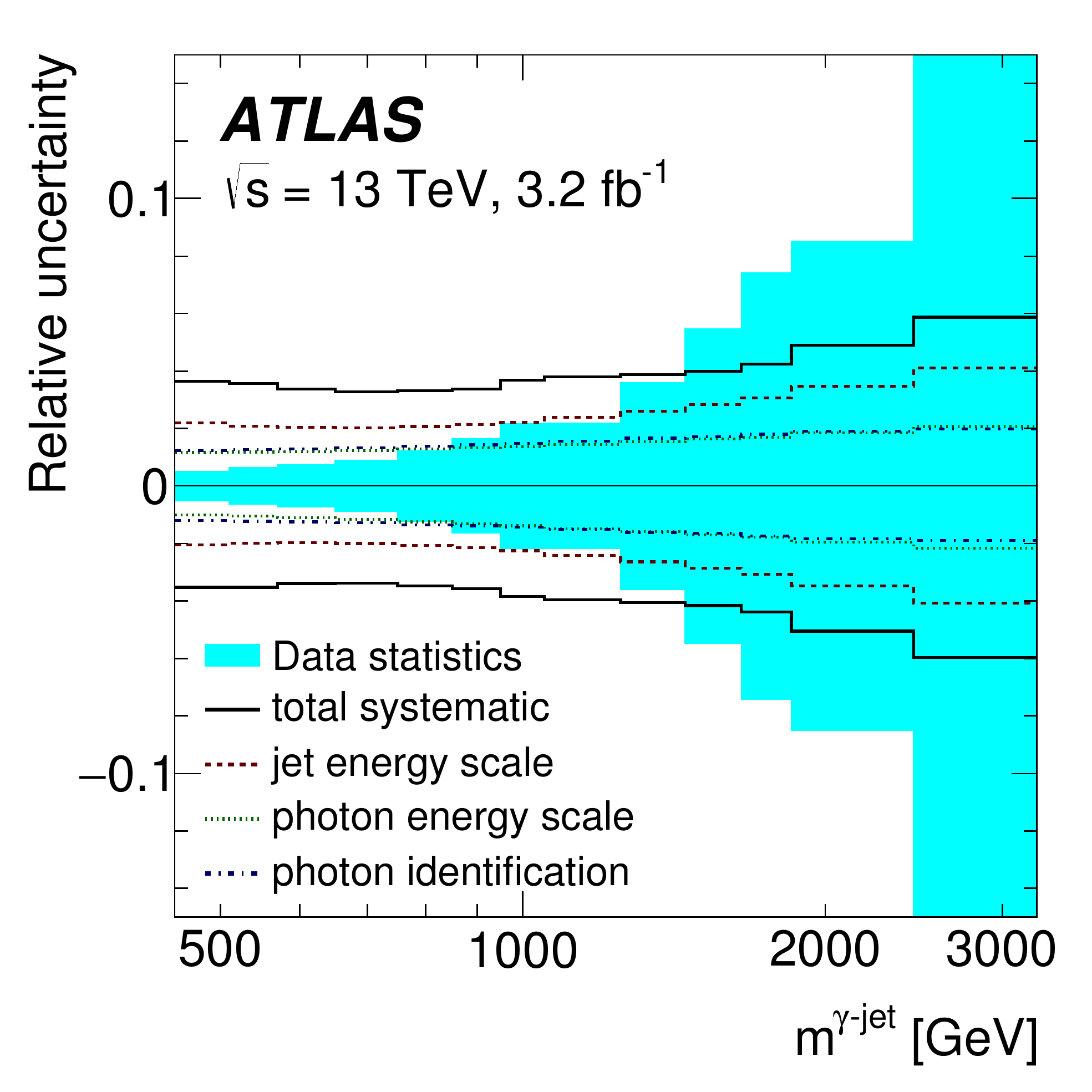}}
\put (8.0,0.0){\includegraphics[width=7cm,height=7cm]{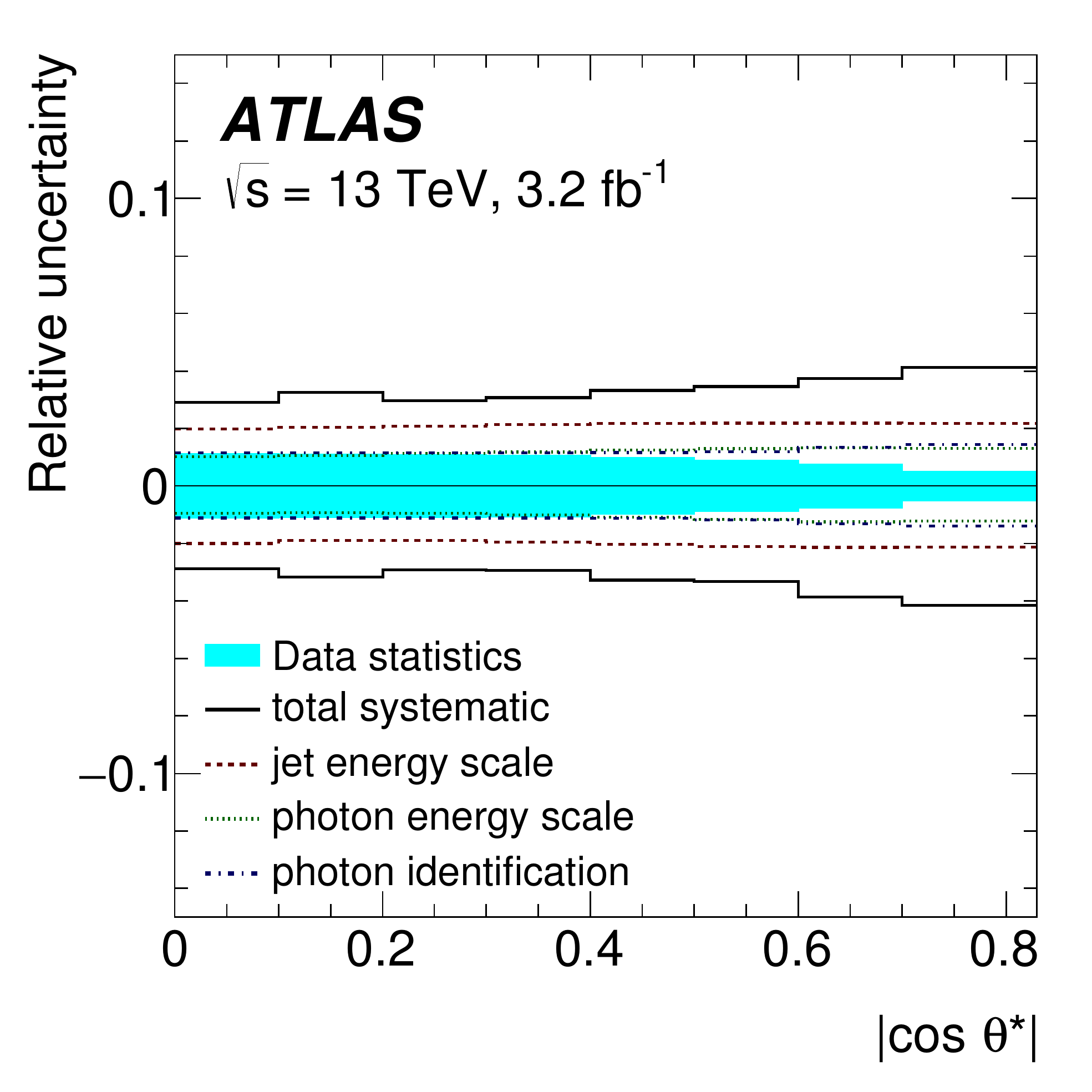}}
\end{picture}
\caption
{
  Total relative systematic uncertainty (solid lines), excluding that
  in the luminosity measurement, as a function of  $\etg$, $\ptjetl$,
  $\deltaphigj$, $\mgj$ and $|\ctgj|$. The three dominant
  contributions are also included separately: the jet energy scale
  (dashed lines), the photon energy scale (dotted lines) and the
  photon identification (dot-dashed lines). The shaded band displays
  the relative statistical uncertainty; for the last point in $\etg\
  (\mgj)$ the relative statistical uncertainty is $32\%$ ($30\%$).
}
\label{fig1}
\end{figure}

The total systematic uncertainty is computed by adding in quadrature
the uncertainties from the sources listed above, the statistical
uncertainty of the MC samples, the uncertainty in the trigger
efficiency ($1\%$) and the uncertainty in the integrated luminosity,
which is fully correlated between all bins of all the measured cross
sections. There are large correlations in the systematic uncertainties
across bins of one observable, particularly in the uncertainties due
to the photon and jet energy scales, which are dominant. The total
systematic uncertainty, excluding that in the luminosity, is less than
$5\%$ for $\etg$, $4\%$ for $\deltaphigj$, $6\%$ for $\mgj$ and $4\%$
for $|\ctgj|$ and increases from $4\%$ at $\ptjetl=100$~\GeV\ to
$10\%$ at $\ptjetl\sim 1.5$~\TeV. Figure~\ref{fig1} shows the total
systematic uncertainty  for each measured cross section, excluding
that in the luminosity; the dominant components are shown separately
in this Figure. The systematic uncertainty dominates the total
experimental uncertainty for $\etg\lesssim 700$~\GeV\ and
$\mgj\lesssim 1.5$~\TeV, whereas for higher $\etg$ and $\mgj$ values,
the statistical uncertainty of the data limits the precision of the
measurements. For $\ptjetl$, $\deltaphigj$ and $|\ctgj|$, the
systematic uncertainty dominates in the whole measured range.

\section{Theoretical predictions}
\label{nlo}
The NLO pQCD predictions presented in this Letter are computed using
two programs, namely \jetp~1.3.1\_2 and \sher~2.2.2. The \jetp\
program includes a full NLO pQCD calculation of both the direct and
fragmentation contributions to the cross section for the $\ppgj$
process. The number of massless quark flavours is set to five. The
renormalisation scale $\muR$, factorisation scale $\muF$ and
\FloatBarrier
fragmentation scale $\muf$ are chosen to be
$\muR=\muF=\muf=\etg$~\cite{pr:d73:094007}. The calculations are
performed using the MMHT2014~\cite{epj:c75:204} PDF set and the BFG
set II of parton-to-photon fragmentation functions~\cite{epj:c2:529},
both at NLO. The strong coupling constant is set to $\asz=0.120$. The
calculations are performed using a parton-level  isolation criterion
which requires the total transverse energy from the partons inside a
cone of size $\Delta R=0.4$ around the photon direction to be below
\etisocutp. The NLO pQCD predictions from \jetp\ are at the parton
level while the measurements are at the particle level. Thus, there
can be differences between the two levels concerning the photon
isolation as well as the photon and jet four-momenta. Since the data
are corrected for pile-up and UE effects and the distributions are
unfolded to a phase-space definition in which the requirement on
$\etiso$ at particle level is applied after subtraction of the UE, it
is expected that parton-to-hadron corrections to the NLO pQCD
predictions are small. Correction factors to the \jetp\ predictions
are estimated by computing the ratio of the particle-level cross
section for a \pyt\ sample with UE effects to the parton-level cross
section without UE effects. These factors are close to unity within
$\pm 5\%$ for the observables studied, except for $\ptjetl\gtrsim
600$~\GeV; in this region, which is dominated by the bremsstrahlung
component, the factors can differ by up to $30\%$ from unity since
hadronisation of a nearby parton can significantly change the
particle-level isolation compared to the parton-level isolation.

The \sher~2.2.2 program consistently combines parton-level
calculations of $\gamma+ {\mathrm{(1,\ 2)-jet}}$ events at NLO and
$\gamma+{\mathrm{(3,\ 4)-jet}}$ events at
LO~\cite{jhep:0202:044,prl:108:111601} supplemented with a parton
shower~\cite{jhep:0803:038} while avoiding double-counting
effects~\cite{jhep:1304:027}. A requirement on the photon isolation at
the matrix-element level is imposed using Frixione's criterion with
${\cal R}=0.1$, $n=2$ and $\epsilon=0.1$. Dynamic factorisation and
renormalisation scales are adopted as well as a dynamical merging
scale with $\bar Q_{\mathrm{cut}}=20$~\GeV~\cite{jp:g44:044007}. The
strong coupling constant is set to $\asz=0.118$. Fragmentation into
hadrons and simulation of the UE are performed using the same models
as for the LO \sher\ samples. The next-to-next-to-leading-order (NNLO)
NNPDF3.0 PDF set~\cite{jhep:1504:040} is used in conjunction with the
corresponding \sher\ tuning. All the NLO \sher\  predictions are based
on the particle-level observables from this computation after applying
the requirements listed in Table~\ref{table1}.

\FloatBarrier
\subsection{Uncertainties in the predictions}
The uncertainty in the NLO pQCD predictions from \jetp\ due to terms
beyond NLO is estimated by repeating the calculations using values of
$\muR$, $\muF$ and $\muf$ scaled by the factors $0.5$ and $2$. The
three scales  are either varied simultaneously, individually or by
fixing one and varying the other two. In all cases, the condition
$0.5\leq\mu_A/\mu_B\leq 2$ is imposed, where $A,B={\mathrm{R,\ F,\
    f}}$. The final uncertainty is taken as the largest deviation from
the nominal value among the 14 possible variations. In the case of the
NLO \sher\ prediction, which does not include the fragmentation
contribution, $\muR$ and $\muF$ are varied as above and the largest
deviation from the nominal value among the 6 possible variations is
taken as the uncertainty.

The uncertainty in the NLO pQCD predictions from \jetp\ due to the
choice of proton PDFs is estimated by repeating the calculations using
the $50$ sets from the MMHT2014 error analysis~\cite{epj:c75:204} and
applying the Hessian method~\cite{pr:d65:014013} for evaluation of the
PDF uncertainties. In the case of NLO \sher, it is estimated using 100
replicas from the NNPDF3.0 analysis~\cite{jhep:1504:040}.

The uncertainty in the NLO pQCD predictions from \jetp\ (NLO \sher)
due to the uncertainty in $\as$ is estimated by repeating the
calculations using two additional sets of proton PDFs from the
MMHT2014 (NNPDF3.0) analysis, for which different values of $\as$ at
$\mz$ were assumed in the fits, namely $0.118$ ($0.117$) and $0.122$
($0.119$); in this way, the correlation between $\as$ and the PDFs is
preserved.

The uncertainty in the parton-to-hadron correction is estimated by
comparing the values obtained using different tunes of \pyt: the ATLAS
set of tuned parameters for the underlying event (tune
AU2)~\cite{ATL-PHYS-PUB-2012-003} with the CTEQ6L1 PDF
set~\cite{jhep:0207:012}, the A14 tune with the LO NNPDF2.3 PDF set as
well as the tunes in which the parameter settings of the latter
related to the modelling of the UE are
varied~\cite{ATL-PHYS-PUB-2014-021}. Larger differences are obtained
from the comparison of the two central tunes than from the variations
around the A14 tune. The nominal correction is taken as the average of
the corrections using the two central tunes, while the uncertainty is
estimated as half of the difference between the two central tunes.

The dominant theoretical uncertainty is that arising from the terms
beyond NLO and, in the case of \jetp\ (NLO \sher), is $\approx 10\%$
($15$--$25\%$) for $\etg$, $\mgj$ and $|\ctgj|$ and increases from
$5\%$ ($15\%$) at $\ptjetl=130$~\GeV\ to $30\%$ ($30\%$) for
$\ptjetl=1.5$~\TeV. In the case of the NLO \sher\ prediction for
$\dsphi$, the uncertainty increases from $10\%$ at
$\deltaphigj\sim\pi$ to $40\%$ at $\deltaphigj\sim\pi/2$. The
uncertainty in the predictions of \jetp\ (NLO \sher) arising from that
in the PDFs is $\lesssim 2\%$ ($3\%$) for all observables. The
uncertainty arising from the value of $\asz$ is below $2\%$ ($5\%$).
The uncertainty in the parton-to-hadron correction is in the range
$1$--$3\%$ except for $\ptjetl\gtrsim 600$~\GeV, where it increases to
$20\%$ at $\ptjetl = 1.5$~\TeV; this uncertainty is included in the
\jetp\ predictions, but not in the case of  NLO \sher\ since it is a
particle-level Monte Carlo generator.\footnote{An uncertainty related
  to the modelling of the hadronisation process should also be
  assigned to the NLO \sher\ predictions, but no tune other than the
  default one is available. It is expected that the uncertainty should
  be of similar size as that evaluated using \pyt.} The total
theoretical uncertainty is obtained by adding in quadrature the
individual uncertainties listed above and, in the case of \jetp\ (NLO
\sher), is $10$--$15\%$ ($15$--$25\%$) except for $\ptjetl$, where it
is in the range $10$--$40\%$ ($15$--$30\%$); in the case of the NLO
\sher\ prediction for $\dsphi$, the total uncertainty is
$10$--$40\%$.

\section{Results}
\label{res}
The measurements presented here apply to isolated prompt photons with
$\etiso<$~\etisocutp\ at particle level and jets of hadrons. The
measured fiducial cross section for isolated-photon plus one-jet
production in the phase-space region given in Table~\ref{table1} is 
$\sigma_{\mathrm{meas}} = 300 \pm 10\ ({\mathrm{exp.}})\ \pm 6\ ({\mathrm{lumi.}})\ {\mathrm{pb}},$
where ``exp.'' denotes the sum in quadrature of the statistical and
systematic uncertainties excluding that due to the luminosity and
``lumi.'' denotes the uncertainty due to that in the integrated
luminosity. The fiducial cross sections predicted by NLO QCD \jetp\
(multi-leg NLO QCD plus parton-shower \sher) using the MMHT2014
(NNPDF3.0) PDF set are

\vspace{0.25cm}
\centerline{$\sigma_{\mathrm{\jetp}}=291_{-21}^{+25}\ ({\mathrm{scale}})\  _{-3}^{+2}\ ({\mathrm{PDF}})\ _{-5}^{+4} \ ({\mathrm{\alpha_{\mathrm{s}}}})\ \pm 6 \ ({\mathrm{non\mhyphen perturb.}})\ {\mathrm {pb}}$}

and

\centerline{$\sigma_{\mathrm{NLO \sher}}=319_{-45}^{+54}\ ({\mathrm{scale}})\ \pm 3\ ({\mathrm{PDF}})\ _{-11}^{+10} \ ({\mathrm{\alpha_{\mathrm{s}}}})\ {\mathrm {pb}},$}
which are consistent with the measurement within the theoretical uncertainties.

\begin{figure}[h]
\setlength{\unitlength}{1.0cm}
\begin{picture} (18.0,20.0)
\put (0.0,13.0){\includegraphics[width=8cm,height=8cm]{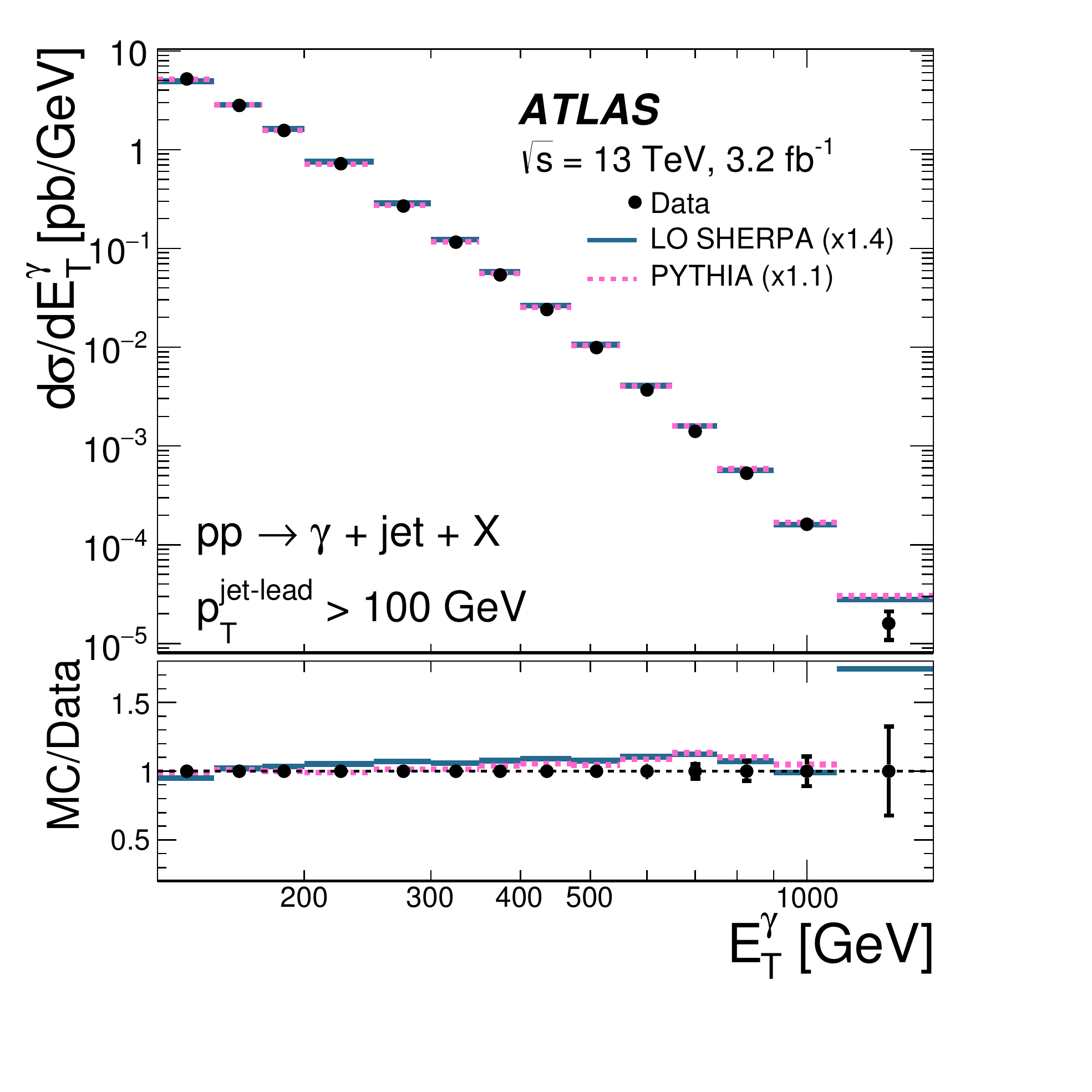}}
\put (8.0,13.0){\includegraphics[width=8cm,height=8cm]{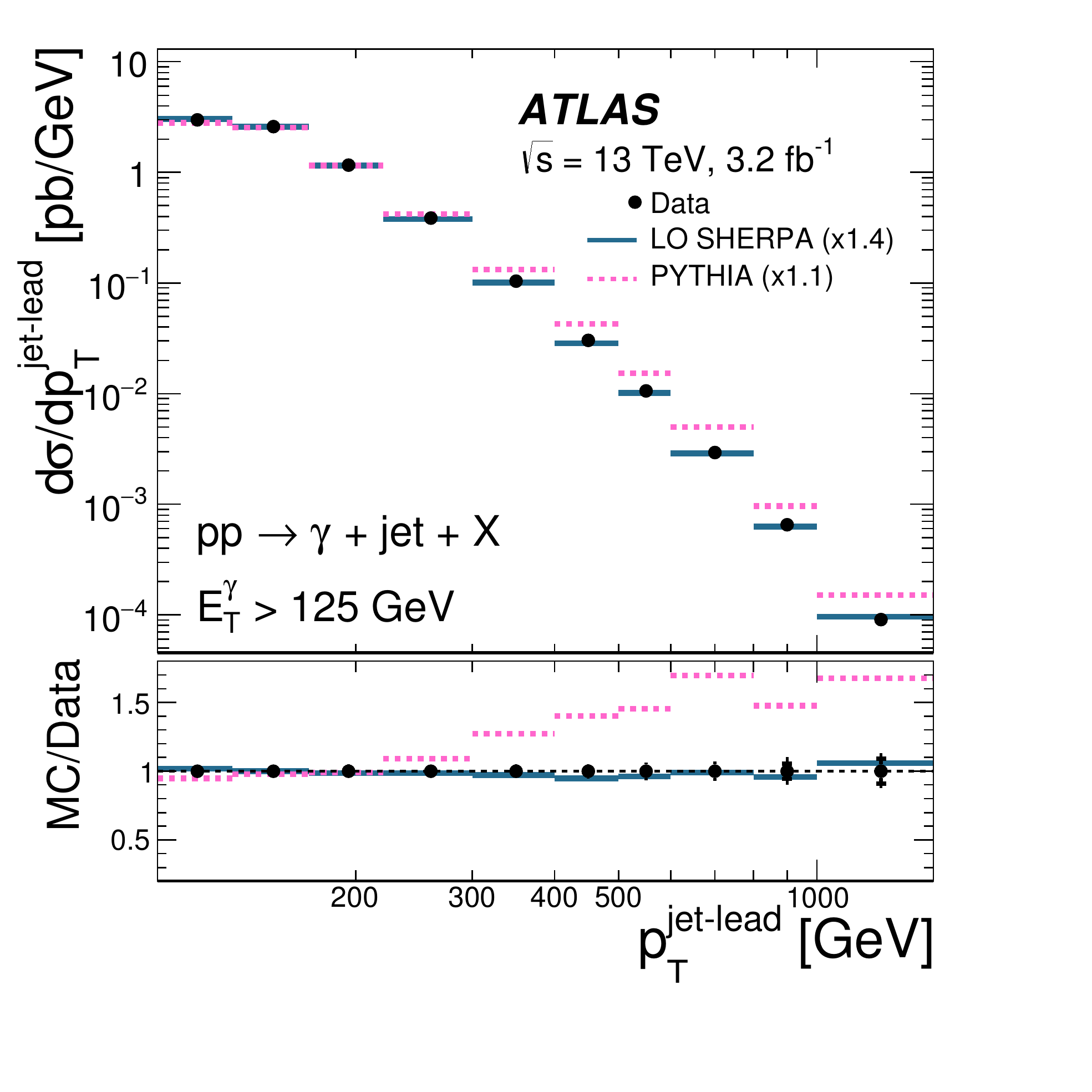}}
\put (0.0,6.0){\includegraphics[width=8cm,height=8cm]{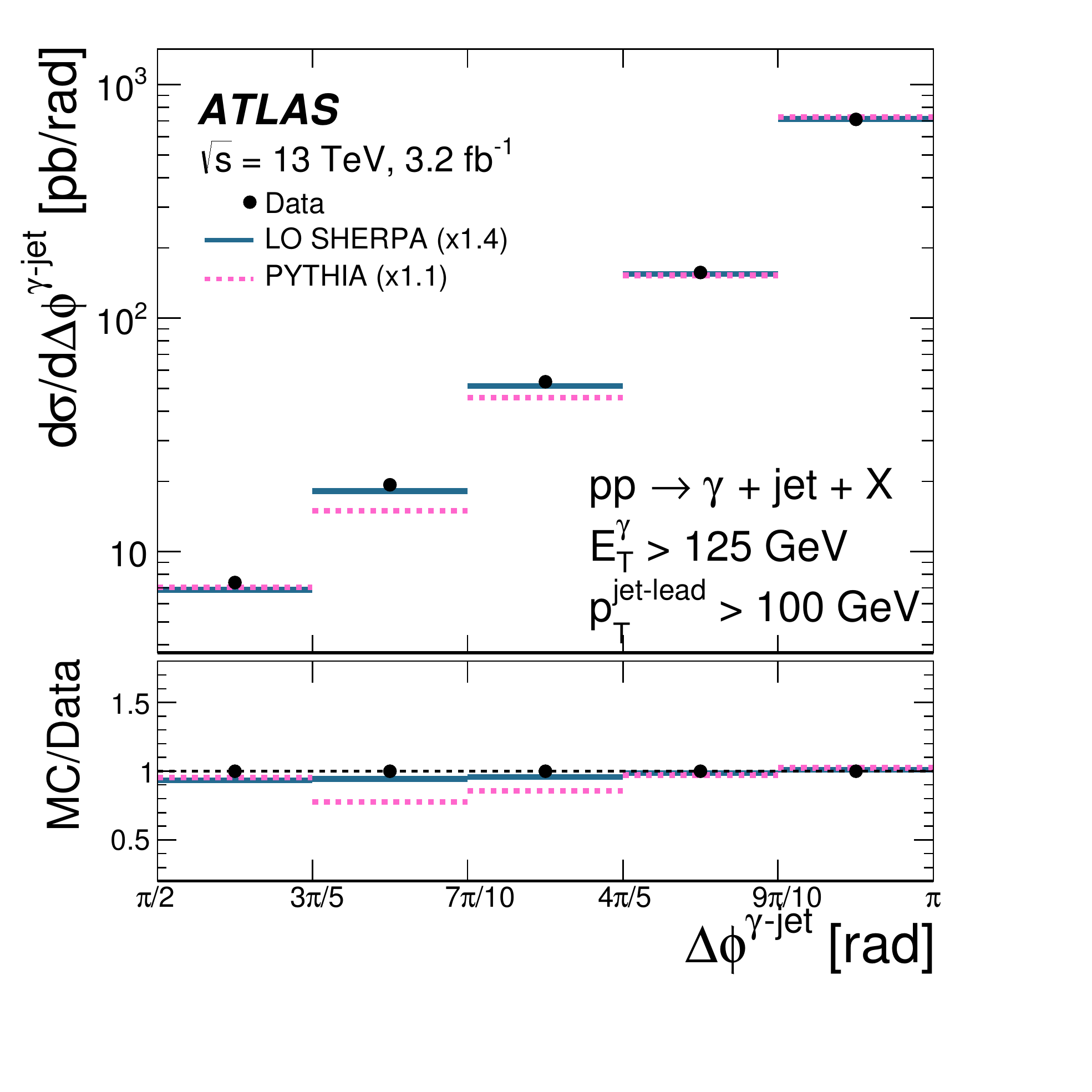}}
\put (0.0,-1.0){\includegraphics[width=8cm,height=8cm]{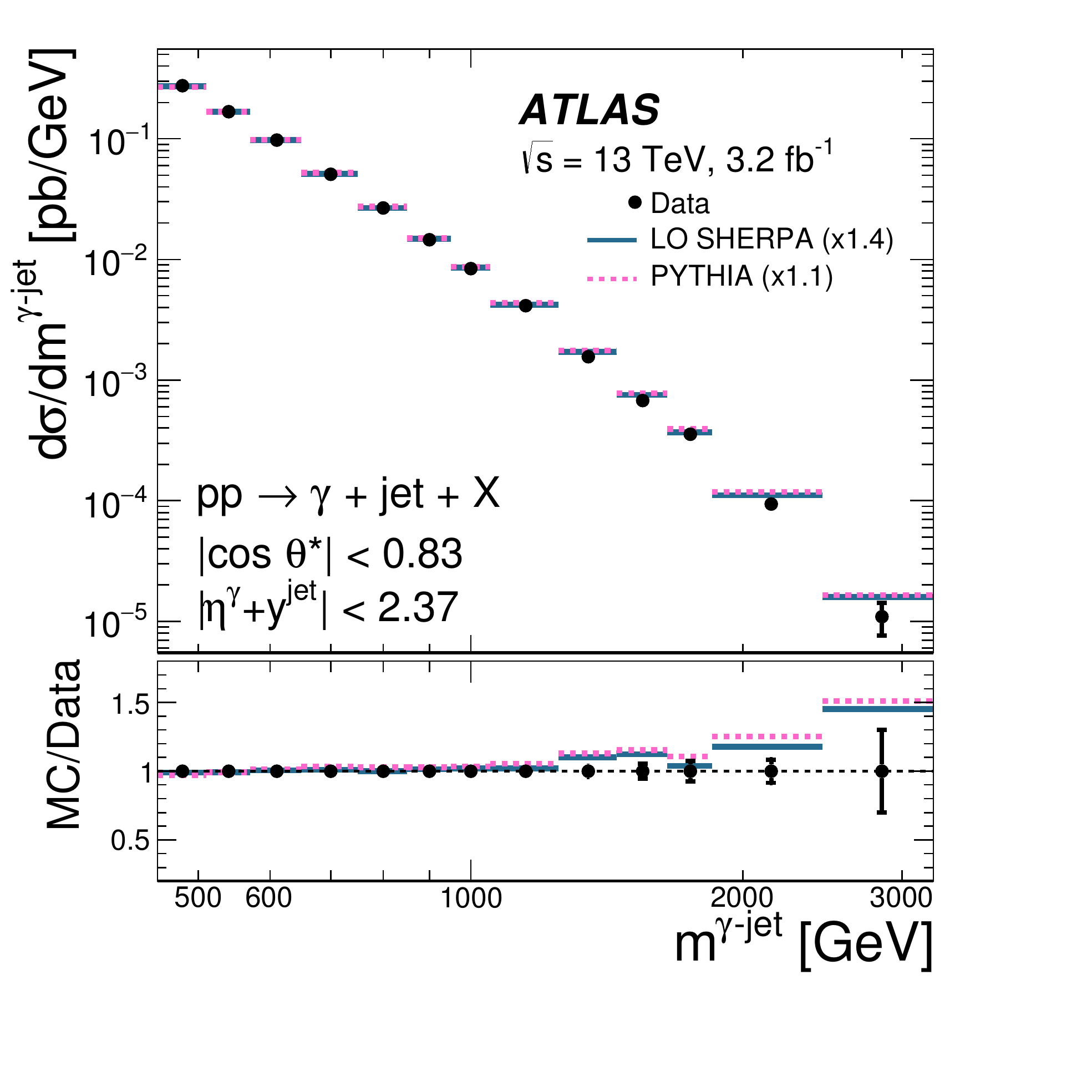}}
\put (8.0,-1.0){\includegraphics[width=8cm,height=8cm]{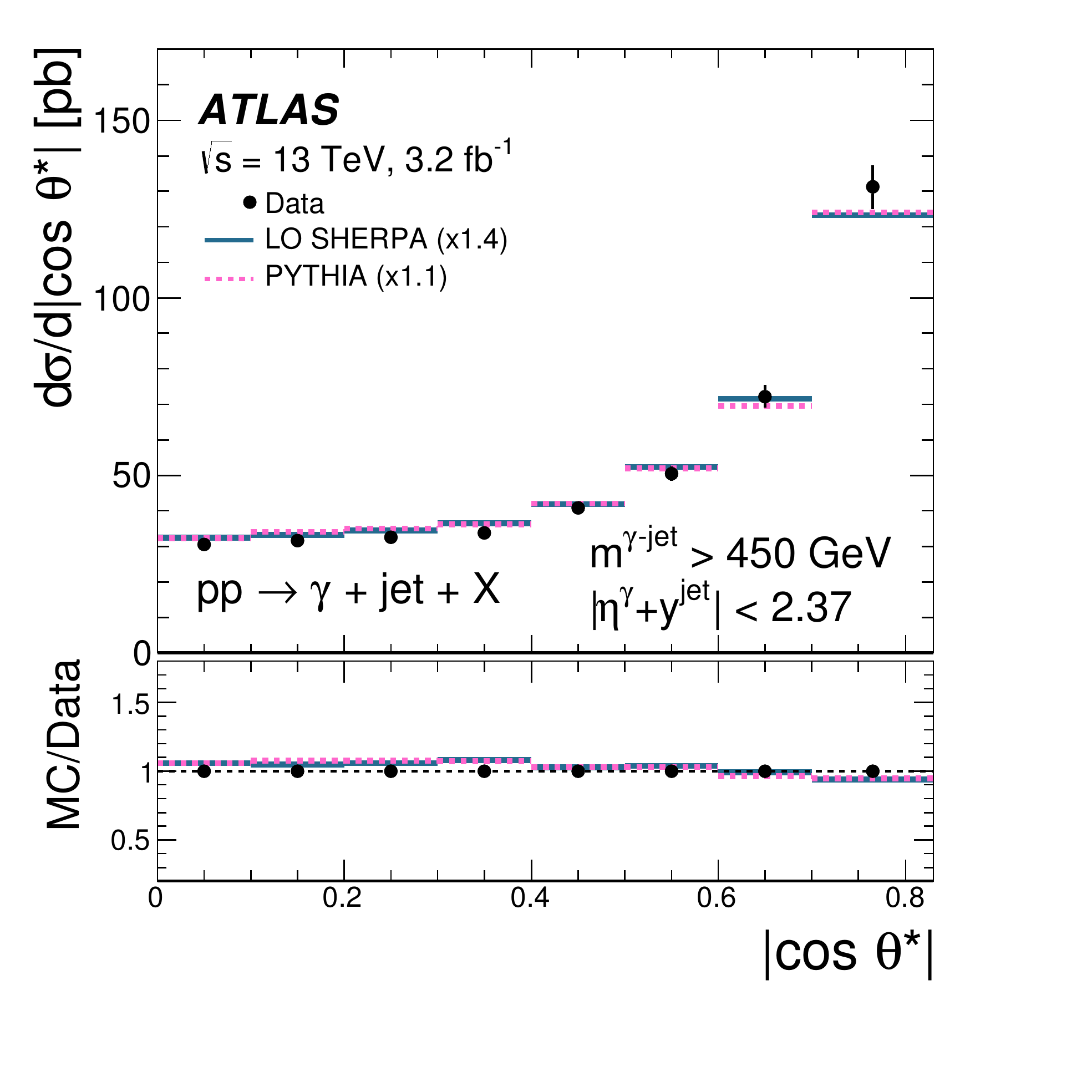}}
\end{picture}
\caption
{
  Measured cross sections for isolated-photon plus jet production
  (dots) as functions of $\etg$, $\ptjetl$, $\deltaphigj$, $\mgj$ and
  $|\ctgj|$; the observables are constructed using the leading photon
  and the leading jet.  For comparison, the tree-level plus
  parton-shower predictions from LO \sher\  (solid lines)  and \pyt\
  (dashed lines) normalised to the integrated measured cross sections
  (using the factors indicated in parentheses) are also shown. The
  theoretical uncertainties associated with the tree-level predictions
  are not included. The bottom part of each figure shows the ratios of
  the MC predictions to the measured cross section. The inner (outer)
  error bars represent the statistical uncertainties (the statistical
  and systematic uncertainties added in quadrature). For most of the
  points, the inner error bars are smaller than the marker size and,
  thus, not visible.
}
\label{fig2}
\end{figure}

Figure~\ref{fig2} shows the isolated-photon plus jet cross sections as
functions of $\etg$, $\ptjetl$, $\deltaphigj$, $\mgj$ and
$|\ctgj|$. The measured $\dsetg$ decreases by almost six orders of
magnitude over the $\etg$ range studied. Values of $\etg$ up to
$1.5$~\TeV\ are measured. The measured $\dspt$ decreases by more than
four orders of magnitude from $\ptjetl = 100$~\GeV\ up to the highest
measured value, $\ptjetl\approx 1.5$~\TeV. The measurement of $\dsphi$
is restricted to $\deltaphigj>\pi/2$ to avoid the phase-space region
dominated by photon production in association with a multi-jet
system. The measured $\dsphi$ increases as
\FloatBarrier
$\deltaphigj$ increases. The measured $\dsmgj$ decreases by more than
four orders of magnitude up to the highest measured value,
$\mgj=3.25$~\TeV. The measured $\dsctgj$ increases as $|\ctgj|$
increases.

The tree-level predictions of the \pyt\ and LO \sher\ MC models are
compared to the measurements in Figure~\ref{fig2}. These predictions
are normalised to the measured integrated fiducial cross section. The
difference in normalisation between data and \pyt\ (LO \sher) is $\sim
+10\%\ (+40\%)$ and attributed to the fact that these generators are
based on tree-level matrix elements, which are affected by a large
normalisation uncertainty due to missing higher-order terms; for this
reason, the theoretical uncertainties are not included in
Figure~\ref{fig2}. Both predictions give an adequate description of
the shape of the measured $\dsetg$, although \pyt\ is slightly better
than LO \sher\ for $\etg\lesssim 600$~\GeV. For $\dspt$, the
prediction from LO \sher\ gives an adequate description of the data in
the whole measured range, whereas that from \pyt\ overestimates the
data for $\ptjetl\gtrsim 200$~\GeV; the overestimation is attributed
to a large contribution from photon bremsstrahlung predicted by the
tune used in \pyt\ (see Section~\ref{datmc}). The prediction from LO
\sher\ gives a good description of the measured $\dsphi$, whereas
\pyt\ underestimates the data for
$3\pi/5<\deltaphigj<4\pi/5$~rad; this is expected from the inclusion
of additional partons in the matrix elements in \sher\ as compared to
\pyt, for which additional partons must necessarily come from the
parton shower. Both predictions give a good description of the data
for $\mgj<1.25$~\TeV\ and for all of the measured $|\ctgj|$ range.

\begin{figure}[h]
\setlength{\unitlength}{1.0cm}
\begin{picture} (18.0,5.5)
\put (0.0,-2.2){\centerline{\includegraphics[width=8cm,height=8cm]{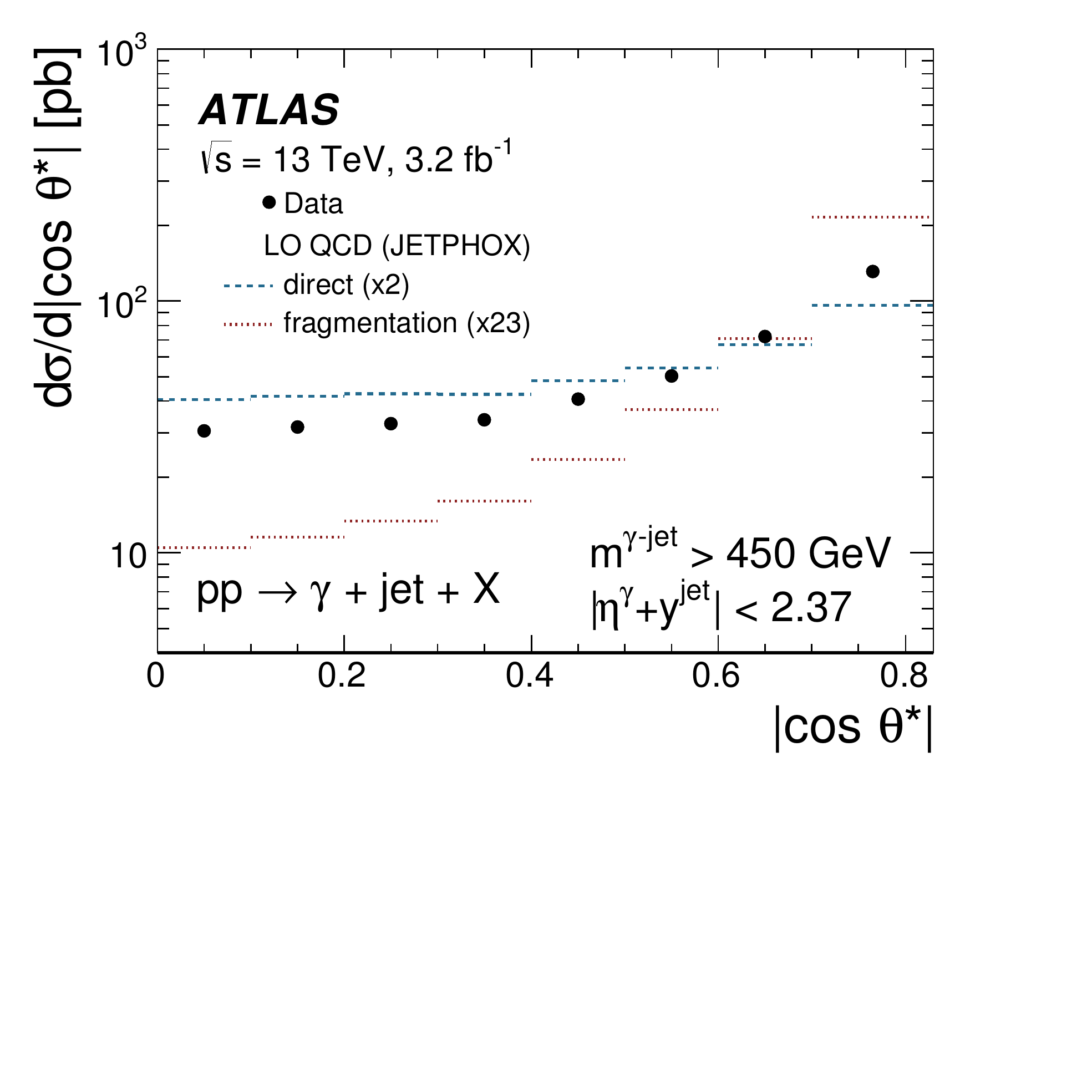}}}
\end{picture}
\vspace{-1.0cm}
\caption
{
  Measured cross section for isolated-photon plus jet production
  (dots) as a function of $|\ctgj|$; the observable is constructed
  using the leading photon and the leading jet. For comparison, the LO
  QCD predictions from \jetp, normalised to the integrated measured
  cross section by the factors shown in parentheses, of direct (dashed
  lines) and fragmentation (dotted lines) processes are shown
  separately. The error bars are smaller than the marker size and,
  thus, not visible.
}
\label{fig5}
\end{figure}

To illustrate the sensitivity to $t$-channel quark or gluon exchange,
the predicted cross-sections $\dsctgj$ from \jetp\ for LO direct and
fragmentation processes are compared to the measurement in
Figure~\ref{fig5}. Even though the two components are no longer
distinguishable at NLO, the LO calculations are useful in illustrating
the basic differences in the dynamics of the two processes. The
contribution from fragmentation, dominated by gluon exchange, shows a
steeper increase as $|\ctgj| \rightarrow 1$ than that from direct
processes, dominated by quark exchange. The shape of the measured
cross-section $\dsctgj$ is closer to that of the direct processes than
that of fragmentation. This is consistent with the dominance of
processes in which the exchanged particle is a quark.

The predictions of the fixed-order NLO QCD calculations of \jetp\
based on the MMHT2014 proton PDF set and corrected for hadronisation
and UE effects as explained in Section~\ref{nlo} are compared to the

\begin{figure}[p]
\setlength{\unitlength}{1.0cm}
\begin{picture} (18.0,20.0)
\put (0.0,13.0){\includegraphics[width=8cm,height=8cm]{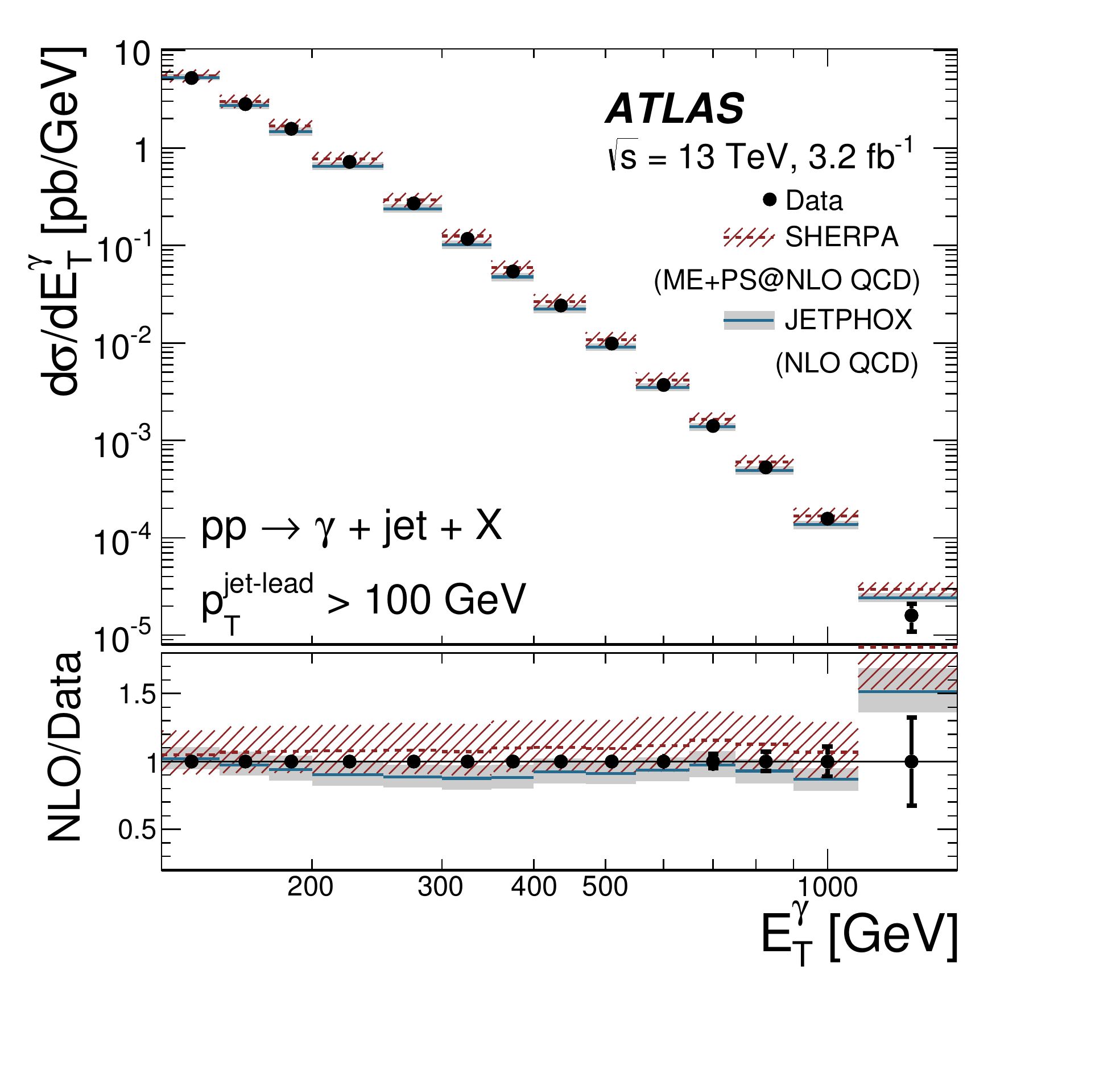}}
\put (8.0,13.0){\includegraphics[width=8cm,height=8cm]{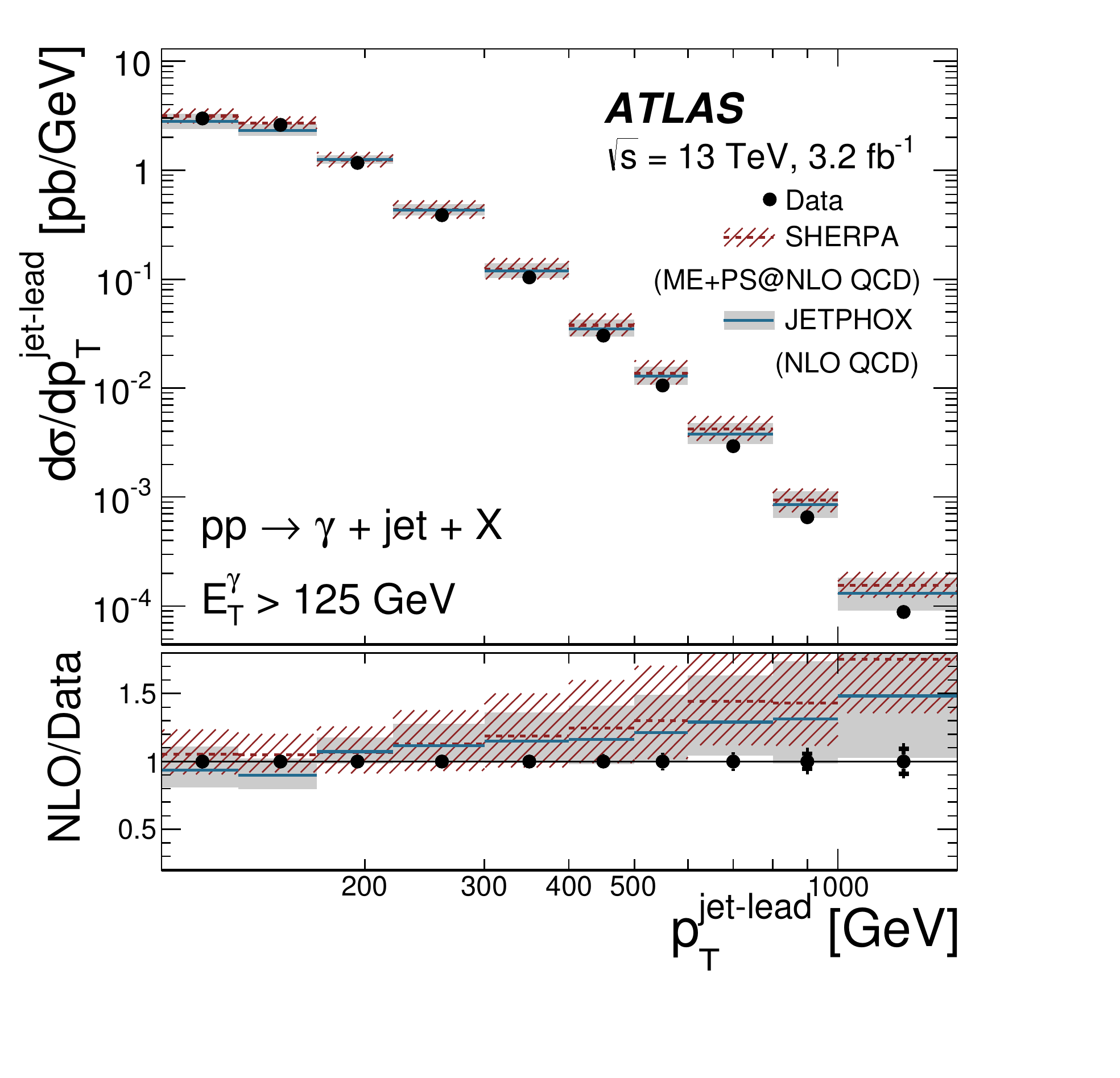}}
\put (0.0,6.1){\includegraphics[width=8cm,height=8cm]{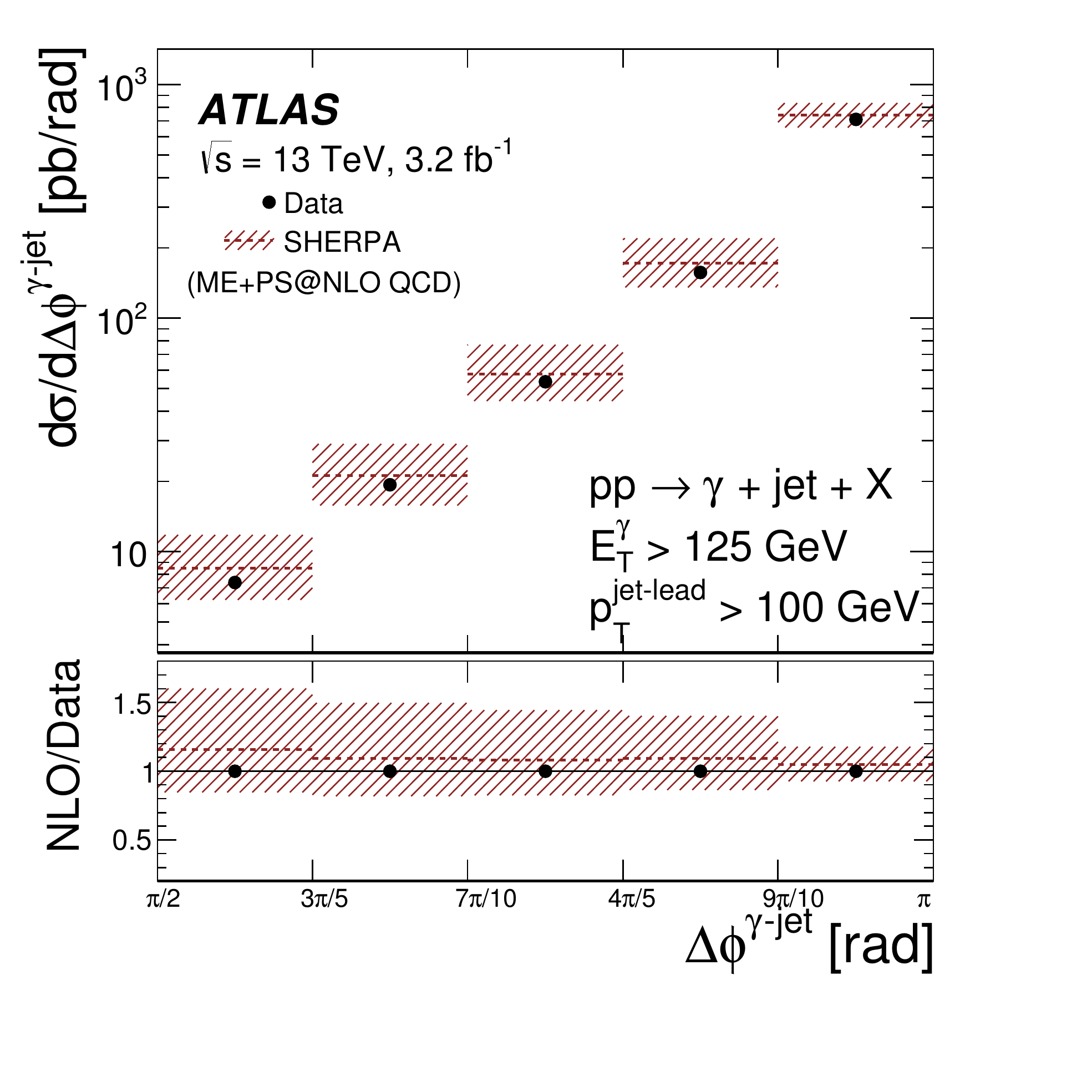}}
\put (0.0,-0.8){\includegraphics[width=8cm,height=8cm]{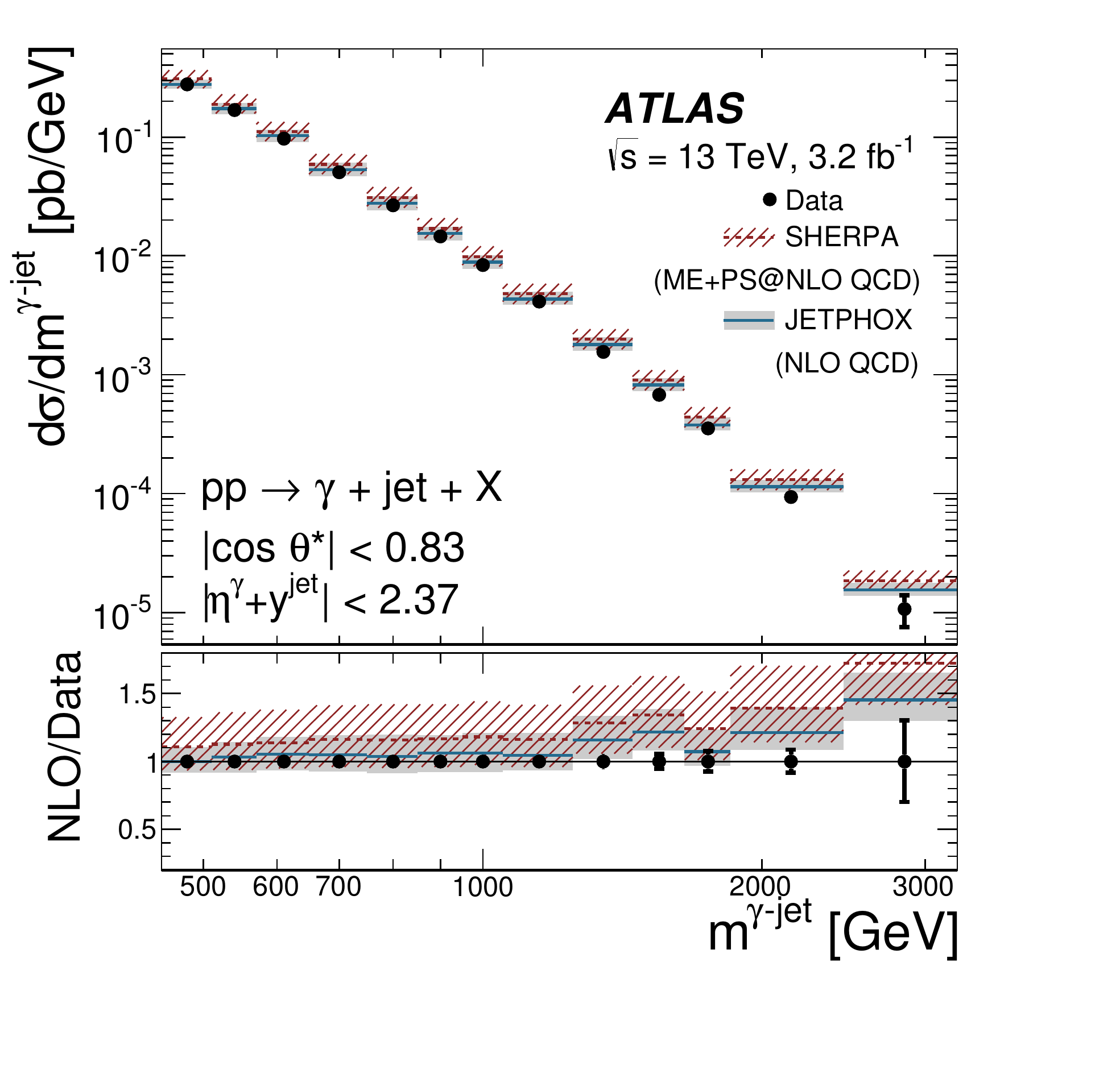}}
\put (8.0,-0.8){\includegraphics[width=8cm,height=8cm]{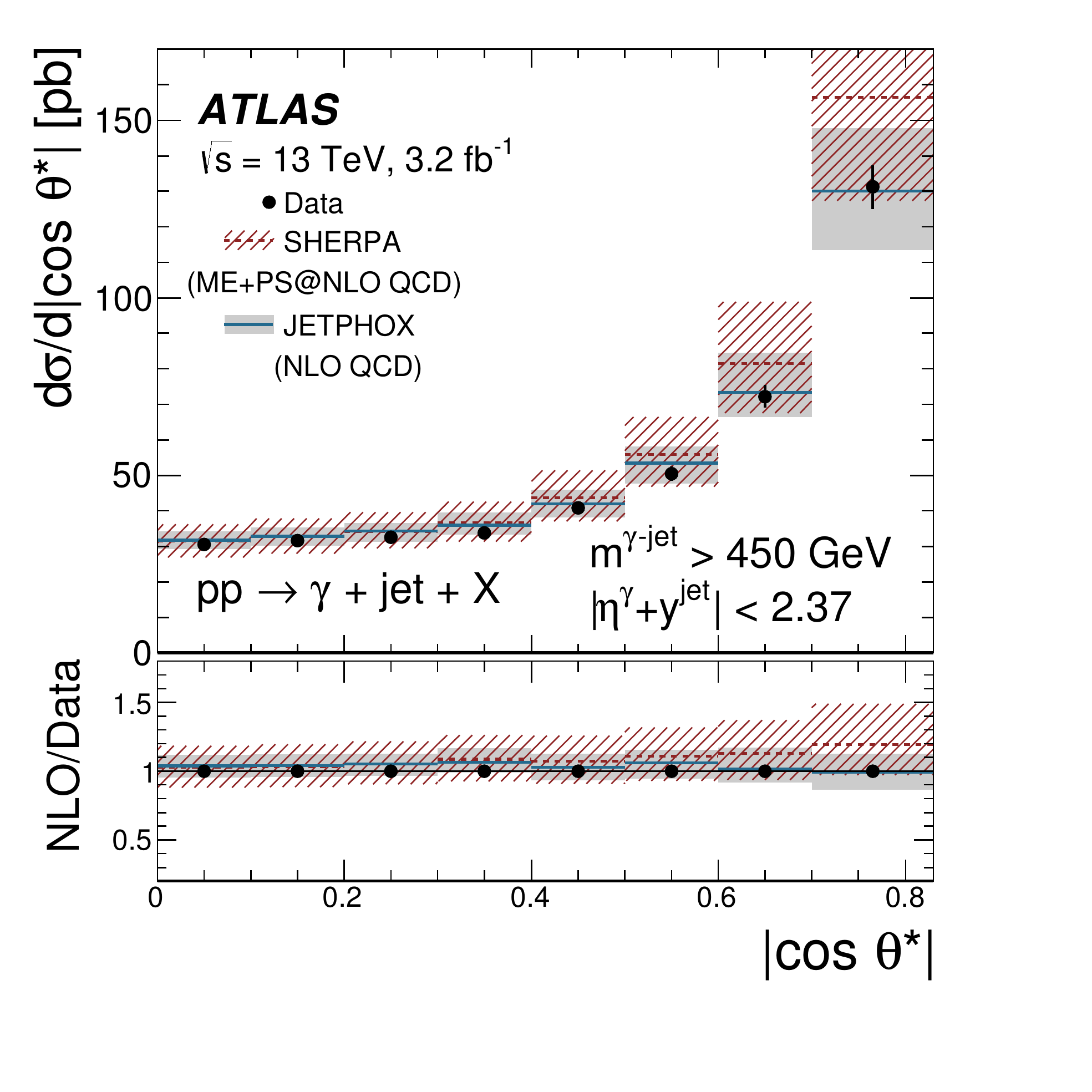}}
\end{picture}
\vspace{-0.7cm}
\caption
{
  Measured cross sections for isolated-photon plus jet production
  (dots) as functions of $\etg$, $\ptjetl$,  $\deltaphigj$,  $\mgj$
  and $|\ctgj|$; the observables are constructed using the leading
  photon and the leading jet.  For comparison, the multi-leg NLO QCD
  plus parton shower predictions from NLO \sher\ (dashed lines) and
  the NLO QCD predictions from \jetp\ corrected for hadronisation and
  UE effects (solid lines) are also shown. The bottom part of each
  figure shows the ratios of the predictions to the measured cross
  section. The inner (outer) error bars represent the statistical
  uncertainties (the statistical and systematic uncertainties added in
  quadrature) and the bands display the theoretical uncertainty. For
  most of the points, the inner error bars are smaller than the marker
  size and, thus, not visible.
}
\label{fig4}
\end{figure}
\FloatBarrier
measurements\footnote{As shown in Ref.~\cite{np:b875:483}, the NLO QCD
  predictions of \jetp\ cannot describe $\dsphi$ due to the limited
  number of final-state partons.} in Figure~\ref{fig4}. The
predictions of the multi-leg NLO QCD plus parton-shower
calculations of \sher\ based on the NNPDF3.0 PDF set are also compared to the
measurements in Figure~\ref{fig4}. Both types of predictions describe
the data within the experimental and theoretical uncertainties. For
the cross section as a function of $\deltaphigj$, the only
well-founded prediction is that of NLO \sher, which is able to
reproduce the data down to $\deltaphigj=\pi/2$ due to the inclusion of
the matrix elements for $2\rightarrow n$ processes with $n=4$ and
$5$. For most of the points, the theoretical uncertainties are larger
than those of experimental origin. Predictions for \jetp\ (\sher\ NLO)
are also obtained with other PDF sets,  namely NLO
CT14~\cite{pr:d93:033006} and NLO NNPDF3.0 (CT14 and MMHT2014),  and
differ by less than $5\%$ with respect to those using MMHT2014
(NNPDF3.0). Thus, the description of the data achieved by the
predictions does not depend significantly on the specific PDF set
used. It is concluded that the NLO pQCD predictions provide an
adequate description of the measurements within the uncertainties.

\section{Summary}
\label{conc}
Measurements of the cross section for the production of an isolated
photon in association with jets in $pp$ collisions at $\sqrt
s=13$~\TeV, $\ppgj$, are presented. These measurements are based on an
integrated luminosity of $3.2$~\fb1\ of ATLAS data recorded at the
LHC. The photon is required to have $\etg>125$~\GeV\ and
$|\etag|<2.37$, excluding the region $1.37<|\etag|<1.56$. The jets are
reconstructed using the anti-$k_t$ algorithm with radius parameter
$R=0.4$. The cross sections are measured as functions of $\etg$,
$\ptjetl$ and $\deltaphigj$ with $\ptjetl>100$~\GeV; the measurements
extend up to values of $1.5$~\TeV\ in $\etg$ and $\ptjetl$. The
dependence on $\mgj$ and $|\ctgj|$ is measured for $\mgj>450$~\GeV.

The predictions of the tree-level plus parton-shower MC models by
\pyt\ and LO \sher\ give a satisfactory description of the shape of
the data distributions, except for $\ptjetl$ in the case of \pyt. The
fixed-order NLO QCD calculations of \jetp, corrected for hadronisation
and UE effects, and the multi-leg NLO QCD plus parton-shower
calculations of \sher\ describe the measured cross sections within the
experimental and theoretical uncertainties. The comparison of
predictions based on different parameterisations of the proton PDFs
shows that the description of the data achieved does not depend
significantly on the specific PDF set used. The only well-founded
prediction for $\dsphi$ is that of NLO \sher, which is able to
reproduce the data down to $\deltaphigj=\pi/2$ due to the inclusion of
the matrix elements for $2\rightarrow n$ processes with $n=4$ and
$5$. The measured dependence on $|\ctgj|$ is consistent with the
dominance of processes in which a quark is exchanged. All these
studies provide tests of the pQCD description of the dynamics of
isolated-photon plus jet production in $pp$ collisions at $\sqrt
s=13$~\TeV. The experimental uncertainties are, in general, much
smaller than the uncertainties in the predictions and, thus,
calculations with higher precision will allow stringent tests of the
theory.

\section*{Acknowledgements}

% Acknowledgements for papers with collision data
% Version 14-Feb-2018

% Standard acknowledgements start here
%----------------------------------------------
We thank CERN for the very successful operation of the LHC, as well as the
support staff from our institutions without whom ATLAS could not be
operated efficiently.

We acknowledge the support of ANPCyT, Argentina; YerPhI, Armenia; ARC, Australia; BMWFW and FWF, Austria; ANAS, Azerbaijan; SSTC, Belarus; CNPq and FAPESP, Brazil; NSERC, NRC and CFI, Canada; CERN; CONICYT, Chile; CAS, MOST and NSFC, China; COLCIENCIAS, Colombia; MSMT CR, MPO CR and VSC CR, Czech Republic; DNRF and DNSRC, Denmark; IN2P3-CNRS, CEA-DRF/IRFU, France; SRNSFG, Georgia; BMBF, HGF, and MPG, Germany; GSRT, Greece; RGC, Hong Kong SAR, China; ISF, I-CORE and Benoziyo Center, Israel; INFN, Italy; MEXT and JSPS, Japan; CNRST, Morocco; NWO, Netherlands; RCN, Norway; MNiSW and NCN, Poland; FCT, Portugal; MNE/IFA, Romania; MES of Russia and NRC KI, Russian Federation; JINR; MESTD, Serbia; MSSR, Slovakia; ARRS and MIZ\v{S}, Slovenia; DST/NRF, South Africa; MINECO, Spain; SRC and Wallenberg Foundation, Sweden; SERI, SNSF and Cantons of Bern and Geneva, Switzerland; MOST, Taiwan; TAEK, Turkey; STFC, United Kingdom; DOE and NSF, United States of America. In addition, individual groups and members have received support from BCKDF, the Canada Council, CANARIE, CRC, Compute Canada, FQRNT, and the Ontario Innovation Trust, Canada; EPLANET, ERC, ERDF, FP7, Horizon 2020 and Marie Sk{\l}odowska-Curie Actions, European Union; Investissements d'Avenir Labex and Idex, ANR, R{\'e}gion Auvergne and Fondation Partager le Savoir, France; DFG and AvH Foundation, Germany; Herakleitos, Thales and Aristeia programmes co-financed by EU-ESF and the Greek NSRF; BSF, GIF and Minerva, Israel; BRF, Norway; CERCA Programme Generalitat de Catalunya, Generalitat Valenciana, Spain; the Royal Society and Leverhulme Trust, United Kingdom.

The crucial computing support from all WLCG partners is acknowledged gratefully, in particular from CERN, the ATLAS Tier-1 facilities at TRIUMF (Canada), NDGF (Denmark, Norway, Sweden), CC-IN2P3 (France), KIT/GridKA (Germany), INFN-CNAF (Italy), NL-T1 (Netherlands), PIC (Spain), ASGC (Taiwan), RAL (UK) and BNL (USA), the Tier-2 facilities worldwide and large non-WLCG resource providers. Major contributors of computing resources are listed in Ref.~\cite{ATL-GEN-PUB-2016-002}.
%----------------------------------------------

\printbibliography

\newpage
% ATLAS Collaboration author list
% Data extracted on 05-Oct-2017 for paper reference STDM-2017-01
\begin{flushleft}
{\Large The ATLAS Collaboration}

\bigskip

M.~Aaboud$^\textrm{\scriptsize 137d}$,
G.~Aad$^\textrm{\scriptsize 88}$,
B.~Abbott$^\textrm{\scriptsize 115}$,
O.~Abdinov$^\textrm{\scriptsize 12}$$^{,*}$,
B.~Abeloos$^\textrm{\scriptsize 119}$,
S.H.~Abidi$^\textrm{\scriptsize 161}$,
O.S.~AbouZeid$^\textrm{\scriptsize 139}$,
N.L.~Abraham$^\textrm{\scriptsize 151}$,
H.~Abramowicz$^\textrm{\scriptsize 155}$,
H.~Abreu$^\textrm{\scriptsize 154}$
% ...
\bigskip
\\
% ...
$^{12}$ Institute of Physics, Azerbaijan Academy of Sciences, Baku, Azerbaijan\\
% ...
$^{88}$ CPPM, Aix-Marseille Universit{\'e} and CNRS/IN2P3, Marseille, France\\
% ...
$^{115}$ Homer L. Dodge Department of Physics and Astronomy, University of Oklahoma, Norman OK, United States of America\\
% ...
$^{119}$ LAL, Univ. Paris-Sud, CNRS/IN2P3, Universit{\'e} Paris-Saclay, Orsay, France\\
% ...
$^{137}$ $^{(a)}$ Facult{\'e} des Sciences Ain Chock, R{\'e}seau Universitaire de Physique des Hautes Energies - Universit{\'e} Hassan II, Casablanca; $^{(b)}$ Centre National de l'Energie des Sciences Techniques Nucleaires, Rabat; $^{(c)}$ Facult{\'e} des Sciences Semlalia, Universit{\'e} Cadi Ayyad, LPHEA-Marrakech; $^{(d)}$ Facult{\'e} des Sciences, Universit{\'e} Mohamed Premier and LPTPM, Oujda; $^{(e)}$ Facult{\'e} des sciences, Universit{\'e} Mohammed V, Rabat, Morocco\\
% ...
$^{139}$ Santa Cruz Institute for Particle Physics, University of California Santa Cruz, Santa Cruz CA, United States of America\\
% ...
$^{151}$ Department of Physics and Astronomy, University of Sussex, Brighton, United Kingdom\\
% ...
$^{154}$ Department of Physics, Technion: Israel Institute of Technology, Haifa, Israel\\
$^{155}$ Raymond and Beverly Sackler School of Physics and Astronomy, Tel Aviv University, Tel Aviv, Israel\\
% ...
$^{161}$ Department of Physics, University of Toronto, Toronto ON, Canada\\
% ...
$^{a}$ Also at Department of Physics, King's College London, London, United Kingdom\\
$^{b}$ Also at Institute of Physics, Azerbaijan Academy of Sciences, Baku, Azerbaijan\\
$^{c}$ Also at Novosibirsk State University, Novosibirsk, Russia\\
$^{d}$ Also at TRIUMF, Vancouver BC, Canada\\
$^{e}$ Also at Department of Physics {\&} Astronomy, University of Louisville, Louisville, KY, United States of America\\
% ...
$^{*}$ Deceased
\end{flushleft}

% Created with xml2latex.py

\end{document}